\documentclass[11pt,a4paper]{article}

\usepackage{amsmath}
\usepackage{amsfonts}
\usepackage{latexsym}

\newcommand{\BEG}[2]{\begin{#1}{#2}\end{#1}}

\newcommand{\GAU}[2]{\MRM{d}\mu_{#1}(#2)}
\newcommand{\PAR}[1]{\frac{\partial}{\partial{#1}}}

\newcommand{\MBB}[1]{{\mathbb{#1}}}

\newcommand{\MCA}[1]{{\mathcal{#1}}}
\newcommand{\MRM}[1]{{\mathrm{#1}}}

\newcommand{\REF}[1]{{(\ref{#1})}}

\begin{document}

\begin{titlepage}
$\phantom{X}$\vspace{2cm}
\begin{center}
{\Large 
Rigorous control of the non-perturbative corrections to the double 
expansion in $g$ and $g^{2}\ln(g)$ for the $\phi^{4}_{3}$-trajectory 
in the hierarchical approximation
\\[10mm]
}
\end{center}
\begin{center}
{\Large C. Wieczerkowski} \\[10mm]
\end{center}
\begin{center}
Institut f\"ur Theoretische Physik I,
Universit\"at M\"unster, \\
Wilhelm-Klemm-Stra\ss e 9, D-48149 M\"unster, \\
wieczer@@uni-muenster.de \\[2mm]
\end{center}
\vspace{-10cm}
\hfill MS-TP1-98-20
\vspace{11cm}
\begin{abstract}\noindent
We study the renormalization invariant trajectory of the 
$\phi^{4}$-perturbation of the free field fixed point in 
the hierarchical approximation. 
We parametrized it by a running $\phi^{4}$-coupling $g$ with 
linear step $\beta$-function.
We rigorously control the non-perturbative corrections to 
finite order approximants from double perturbation theory 
in $g$ and $g^{2}\ln(g)$. 
The construction uses a contraction mapping for the extended 
renormalization group composed of a hierarchical block spin
transformation with a flow of $g$.
\end{abstract}
\end{titlepage}

\section{Introduction}

The non-perturbative renormalization of $\phi^{4}$-theory is 
a central problem of constructive quantum field theory
\cite{Glimm/Jaffe:1987,Rivasseau:1991,Brydges:1992}.
The state of the art includes phase cell expansions 
\cite{Glimm/Jaffe:1973,Feldman/Osterwalder:1976,
Magnen/Seneor:1977,Battle/Federbush:1983,
Feldman/etal:1987},
renormalization group techniques 
\cite{Gallavotti:1978,Benfatto/etal:1980,Balaban:1983,
Gawedzki/Kupiainen:1984,Gallavotti:1985,
Gallavotti/Nicolo:1985,Gawedzki/Kupiainen:1985b,
Pordt:1990}, 
and random path representations
\cite{Brydges/Frohlich/Sokal:1983,Fernandez/Frohlich/Sokal:1992}. 
However, the non-perturbative renormalization of $\phi^4$-theory  
remains a mathematical enterprise of considerable difficulty.
Recent work by Brydges, Dimock, and Hurd 
\cite{Brydges:1994,Brydges:1995,Brydges:1997}
aims to simplify the older constructions and to cast
renormalization theory into a more conceptual form.
The present paper intends to make a modest contribution in 
the same direction. 

A key to the understanding of renormalization theory is 
Wilson's renormalization group
\cite{Wilson:1971,Wilson/Kogut:1974}. 
The traditional starting point is a bare action 
$S_{0}(\phi,g_{0})$\footnote{The quantum field is understood 
to be rescaled to a unit ultraviolet cutoff.}.
The goal is to compute a renormalized action as the limit 
$n\rightarrow\infty$ of 
$S_{n}(\phi,g_{n})=R_{L}^{n}(S_{0})(\phi,g_{0})$ 
of an iterated renormalization group transformation $R_{L}$
with scale $L$. The bare couplings $g_{0}$ are tuned in this 
process so as to obtain a finite limit for $g_{n}$.
This process can be viewed as a trajectory in the dynamical system 
(on some space of actions) generated by $R_{L}$. 
If the action is a fixed point $S_{\star}(\phi)$ of $R_{L}$ then 
its renormalization becomes trivial: the bare action and
the renormalized action become identical. 
This fixed point problem has a natural generalization.
Consider a curve $S(\phi,g)$ of actions parametrized 
by a (running) coupling $g$ such that
\begin{enumerate}
\item $S(\phi,0)$ is a fixed point $S_{\star}(\phi)$ of $R_{L}$,
\item $\partial_{g}S(\phi,g)\vert_{g=0}$ is an eigenvector 
$\MCA{O}(\phi)$ of the linearization of $R_{L}$ at 
$S_{\star}(\phi)$, and 
\item $R_{L}(S)(\phi,g)=S(\phi,\delta^{-1}_{L}(g))$, where 
$\delta_{L}$ is a step $\beta$-function. 
\end{enumerate}
In this case, the bare action and the renormalized action have 
the same functional dependence of $\phi$ but correspond to different 
values of the running coupling $g$. Renormalization amounts to 
control the flow generated by the step $\beta$-function, a 
comparatively easy task. 

We will study a variant of this renormalization problem, where
$R_{L}$ is a block spin transformation for a three dimensional
scalar lattice field theory with hierarchical covariance 
\cite{Koch/Wittwer:1988}. It is designed such that the 
interaction remains local under the renormalization group 
evolution. The lattice interaction Boltzmann factor factorizes
into product of local Boltzmann factors 
$Z(\phi)=\exp\bigl(-V(\phi)\bigr)$ (one for each lattice site), 
which are functions of a real variable $\phi$.
In three dimensions, the hierarchical renormalization group 
then reduces to the non-linear integral transformation
\BEG{gather}{
R_{L}(Z)(\psi)\;=\;
\left\{
   \int\MRM{d}\mu(\zeta)\;
   Z\left(
      L^{-\frac{1}{2}}\psi+\zeta
   \right)
\right\}^{L^{3}},
\label{intro.1}}
where $\MRM{d}\mu(\zeta)$ is the Gaussian measure on $\MBB{R}$
with mean zero and unit covariance. This transformation and 
variants of it have been studied by many authors both as a model 
of constructive renormalization and also because of its properties 
as a non-linear theory. Rigorous work on hierarchical models 
(more generally ultralocal renormalization groups) includes:
\begin{enumerate} 
\item the $\epsilon$-expansion
   \cite{Collet/Eckmann:1977}
\item the $\phi^{4}_{3}$ infrared fixed point 
   \cite{Koch/Wittwer:1986,Koch/Wittwer:1991,Koch/Wittwer:1994}
\item the massive perturbation of the $\phi^{4}_{3}$ fixed point 
   \cite{Koch/Wittwer:1988}
\item the renormalization group differential equation
   \cite{Felder:1987}
\item the $\phi^{4}_{4}$ infrared problem
   \cite{Pordt:1990,Albuquerque:1991}
\item the $\phi^{4}_{4}$ ultraviolet problem at negative coupling
   \cite{Gawedzki/Kupiainen:1985a}
\item the renormalized $\phi^{4}_{D}$-trajectory
   \cite{Wieczerkowski:1997} 
\item the $SU(2)$-lattice gauge theory 
   \cite{Timme:1989}
\item the non-linear $\sigma$-model
   \cite{Gawedzki/Kupiainen:1986,Pordt/Reisz:1991}
\item the sine-Gordon model
   \cite{Dimock:1989,Kappeler/Pinn/Wieczerkowski:1991}
\item multigrid expansions
   \cite{Pordt:1990,Pordt:1993}
\item and random surfaces 
   \cite{Cassandro/Mitter:1994}
\end{enumerate}
Beyond the hierarchical approximation, one has to deal with 
non-local interactions generated by the renormalization group.
Although non-local corrections are rather small in all models
brought under control so far, the mathematical apparatus needed
to control them is a lot more sophisticated, the main tool
being polymer expansions. The virtue of hierarchical models is
that they allow to study renormalization effects without this 
additional burden (or perhaps joy).

In this paper, we continue the work started in 
\cite{Wieczerkowski:1997}. We look for a curve of renormalized
interaction Boltzmann factors $Z(\phi,g)$ with the following 
properties:
\begin{enumerate}
\item $Z(\phi,g)\,=\,\exp\bigl(-g\,:\phi^{4}:\bigr)\;
\biggl(1+O(g^{2}\biggr)$, i.e., $Z(\phi,g)$ emerges from 
the trivial fixed point $Z_{\star}(\phi)=1$ (the free massless
hierarchical field) tangent to a (normal ordered) $\phi^{4}$-vertex,
and
\item $R_{L}(Z)(\psi,\delta_{L}(g))\,=\,Z(\psi,g)$, i.e., 
$Z(\phi,g)$ is a fixed point of the extended renormalization group
$R_{L}\times\delta_{L}^{\star}$ with a linear step $\beta$-function
$\delta(g)=L^{-1}g$. 
\end{enumerate}
The problem is thus to construct a non-trivial fixed point of the
extended renormalization group $S_{L}=R_{L}\times\delta_{L}^{\star}$.
We do this by means of a contraction mapping. For this purpose, 
we split $Z=Z_{1}+Z_{2}$, where $Z_{1}$ is an approximate fixed point, 
and where $Z_{2}$ is a correction. We iterate
the transformation of $Z_{2}$ with $Z_{1}$ kept fixed. This 
transformation is shown to contract, provided that $Z_{1}$ is in
a certain sense a sufficiently good approximation. We compute 
$Z_{1}$ as a polynomial approximant of finite order in a (formal) 
double perturbation expansion in $g$ and $g^{2}\ln(g)$. 
We then prove that this approximant is indeed sufficiently accurate 
provided that the order of perturbation theory is at least seven. 
The result of this construction is the following
Theorem.\footnote{$Z_{QU}$ and $V^{(r_{\MRM{max}})}$ are 
explained in the bulk of this paper.}
\BEG{THE}{
Let $F_{2}(g)$ be a continuous positive function of the form
\BEG{equation}{
F_{2}(g)\;=\;
g^{\sigma}\;
\exp\bigl(
   c_{\star}\,g^{3}+c\,g
\bigr)
\label{intro.2}}
with positive constants $\sigma$, $c_{\star}$, and $c$. Let 
$Z_{QU}(\phi,g)$ be the quadratic fixed point 
\BEG{equation}{
Z_{QU}(\phi,g)\;=\;
\exp\left(
   a_{QU}(g)
   -\frac{b_{QU}(g)}{2}\,\phi^{2}
\right)
\label{intro.3}}
of $S_{L}$. Let $\MCA{B}_{2}$ be the Banach space of functions
$Z_{2}:\MBB{R}\times[0,g_{\MRM{max}}]\rightarrow\MBB{R}$ with
respect to the norm
\BEG{equation}{
\|Z_{2}\|_{F_{2}}\;=\;
\sup_{(\phi,g)\in\MBB{R}\times[0,g_{\MRM{max}}]}\;
\left\vert
   \frac{Z_{2}(\phi,g)}{Z_{QU}(\phi,g)\,F_{2}(g)}
\right\vert.
\label{intro.4}}
Let $Z_{1}(\phi,g)\;=\;\exp\left(-V^{(r_{\MRM{max}})}(\phi,g)
\right)$, where $V^{(r_{\MRM{max}})}(\phi,g)$ is the polynomial 
approximant of order $r_{\MRM{max}}$ (in $g$) of the 
perturbative solution to the fixed point equation as a 
double expansion in $g$ and $g^{2}\ln(g)$. 

For $r_{\MRM{max}}=7$, there exist positive constants $g_{\MRM{max}}$,
$\sigma$, $c_{\star}$, $c$, and $C_{2}$ such that the transformation 
$S_{L}\left(Z_{1}+Z_{2}\right)-Z_{1}$ is a contraction mapping 
on the ball 
\BEG{equation}{
\bigl\{Z_{2}\in\MCA{B}_{2}\big\vert\;\|Z_{2}\|_{F_{2}}\leq C_{2}\bigr\}.
\label{intro.5}}
} 
It follows that there exists a unique fixed point in this ball. 
Furthermore, the iteration of the contraction mapping gives a 
convergent representation for this fixed point. 

The renormalized $\phi^{4}_{D}$-trajectory was constructed in 
\cite{Wieczerkowski:1997} for all dimensions $2<D<4$ with the 
exception of a discrete set of special dimensions, 
where resonances of power counting factors occur 
\cite{Rolf/Wieczerkowski:1996}. Unfortunately, the case $D=3$
is such a resonant case, and was therefore excluded in 
\cite{Wieczerkowski:1997}. The problem is that our 
renormalization problem does not have a formal power series
solution in $g$ in three dimensions. However, it does
have a solution as a formal double perturbation expansion 
in $g$ and $g^{2}\ln(g)$. The main content of this paper is 
to deal with these logarithmic corrections. Polynomial 
approximants from this double perturbation theory turn out
to suffice for the contraction mapping. The backbone of our 
approach is the contraction mapping. This part is identical in
resonant and non-resonant dimensions. To keep this paper 
selfcontained we have included a section on the contraction
mapping with fresh proofs and improved bounds as compared to
\cite{Wieczerkowski:1997}. In particular, we present
\begin{enumerate}
\item a better scheme independent proof of the contraction property, 
\item an example of bounds, which are true for all couplings and
      not only small couplings,
\item a better and more explicit treatment of the tree approximation,
\item and last not least a full stability analysis of the 
      $g$-$g^{2}\ln(g)$-approximants.
\end{enumerate}
Unlike \cite{Wieczerkowski:1997}, analyticity in $\phi$ is not
used here. This paper is organized as follows. Section two
contains a brief review of the hierarchical renormalization 
group. Section three is devoted to the contraction mapping 
method. In Section four, we prove a stability bound and an 
error bound for the first order approximant. It serves as a 
template for the higher order approximants, which are 
analyzed in Section five. We conclude with a few remarks and
outlooks.

\section{Hierarchical renormalization group}

The hierarchical renormalization group is the theory of a non-linear
integral transformation $R$ acting on a certain space of functions 
$Z:\MBB{R}\rightarrow\MBB{R}$. We speak of $Z$ as a hierarchical 
model in Statistical Physics if $Z$ is positive, continuous, and 
in a certain sense measurable. 

\subsection{Renormalization group transformation}

Hierarchical renormalization groups come in different versions. 
We mention the transformations of Dyson \cite{Dyson:1969}, 
Wilson \cite{Wilson:1971}, Baker \cite{Baker:1972},and Gallavotti 
\cite{Gallavotti:1978}. The latter version has been investigated 
as a model for the constructive renormalization of massless scalar 
field theories by Gawedzki and Kupiainen 
\cite{Gawedzki/Kupiainen:1984,Gawedzki/Kupiainen:1985a}, by 
Pordt \cite{Pordt:1990,Pordt:1993}, and by
Koch and Wittwer \cite{Koch/Wittwer:1986,Koch/Wittwer:1991}. 
Following their path, we define $R$ as follows.
\BEG{DEF}{
Let $L$ denote a positive integer, $L\in\{2,3,4,\ldots\}$, and 
let $D=3$. Let $\alpha\,=\,L^{D}$, $\beta\,=\,L^{1-\frac{D}{2}}$, 
and $\gamma\,=\,1$. Let $R$ be the non-linear integral transformation 
given by
\BEG{gather}{
R(Z)(\psi)\;=\;
\left\{
   \int\MRM{d}\mu_{\gamma}(\zeta)\,Z(\beta\psi+\zeta)
\right\}^{\alpha},
\label{model.1}}
where $\MRM{d}\mu_{\gamma}(\zeta)$ denotes the Gaussian 
measure\footnote{The Fourier transform of the Gaussian measure is 
$\int\GAU{\gamma}{\zeta}\,\MRM{e}^{\MRM{i}\zeta\psi}\,=\,
\MRM{e}^{-\gamma\psi^{2}/2}$.} on $\MBB{R}$ with mean zero and 
covariance $\gamma$.
}
$L$ is the block scale. Eventually, we will choose $L$ to be large. 
$D$ is the dimension of the (Euclidean) space-time. We will restrict 
our attention to the three dimensional transformation. In the 
sequel, $D$ and $\gamma$ are constant, $D=3$ and $\gamma=1$.
The transformation \REF{model.1} calls to be supplemented by a domain. 
The following statement defines our model space. 
\BEG{DEF}{
Let $\MCA{M}$ be the space of continuous even functions 
$Z:\MBB{R}\rightarrow\MBB{R}$ with finite norm $\|Z\|_{\infty}\,=\,
\sup_{\phi\in\MBB{R}}\vert Z(\phi)\vert$.}
A part of our analysis will be to restrict $R$ to suitable 
invariant subspaces of $\MCA{M}$, for instance subspaces 
of functions with a rapid decrease at infinity. Positive 
functions form a subset $\MCA{M}^{+}$ of $\MCA{M}$. 
But $\MCA{M}^{+}$ is not a linear subspace of $\MCA{M}$. Because we 
intend to use Banach space theory, we take the latter as our 
starting point.
\BEG{LEM}{
The transformation $R$ acts on the space $\MCA{M}$. It satisfies the 
bound $\|R(Z)\|_{\infty}\,\leq\,\|Z\|_{\infty}^{\alpha}$.
}
\subsection{Trivial fixed point and tangent map}

$R$ has two trivial non-zero fixed points in $\MCA{M}^{+}$, the 
ultraviolet fixed point $Z_{UV}(\phi)\,=1\,$, and the (Gaussian) 
high temperature fixed point 
\BEG{gather}{
Z_{HT}(\phi)\;=\;
A_{HT}\;\MRM{e}^{-\frac{b_{HT}}{2}\,\phi^2},
\label{model.2}}
where 
\BEG{gather}{
A_{HT}\;=\;
(\alpha\beta^{2})^{\frac{\alpha}{2(\alpha-1)}}\;=\;
L^{\frac{1}{1-L^{-D}}},
\qquad
b_{HT}\;=\;
\frac{\alpha\beta^{2}-1}{\gamma}\;=\;
L^{2}-1.
\label{model.3}}
In this paper, we will study the unstable manifold of the 
ultraviolet fixed point. Basic knowledge about it is gained 
from the linearized renormalization group. Recall the following 
well known facts.  
\BEG{LEM}{
Let $\MCA{O}\in\MBB{R}[\phi^{2}]$ be an even polynomial in $\phi$ 
with coefficients in $\MBB{R}$ of the form 
$\MCA{O}(\phi)=\phi^{2n}+\text{lower powers of }\phi^{2}$. 
Let $Z:\MBB{R}\times\MBB{R}^{+}
\rightarrow\MBB{R}$ be defined by 
\BEG{gather}{
Z(\phi,g)\;=\;
\MRM{e}^{-g\,\MCA{O}(\phi)}.
\label{model.4}}
(I) For all $g\in\MBB{R}^{+}$, $Z(\cdot,g)\in\MCA{M}^{+}$, and
$Z(\phi,0)\,=\,Z_{UV}(\phi)$.
(II) For all $g\in\MBB{R}^{+}$, $R(Z)(\cdot,g)$ is continuously
differentiable in $g$, and $\lim_{g\rightarrow 0^{+}}\partial_{g}
R(Z)(\cdot,g)$ defines a linear operator
\BEG{gather}{
D_{Z_{UV}}(R)(\MCA{O})(\psi)\,=\,
\alpha\int\GAU{\gamma}{\zeta}\,\MCA{O}(\beta\psi+\zeta).
\label{model.5}}
on $\MBB{R}[\phi^{2}]$.
}
The linear operator $D_{Z_{UV}}(R)$ is the tangent map of $R$ at
$Z_{UV}$. This tangent map is diagonalizable. 
\BEG{LEM}{
Let $\MRM{H}_{2n}$ be the Hermite polynomial of order $2n$ 
\cite[8.95]{Gradshteyn/Ryzhik:1980}.
Let $P_{2n}\in\MBB{R}[\phi^{2}]$ be given by
\BEG{gather}{
P_{2n}(\phi,\upsilon)\;=\;
\left(\frac{\upsilon}{2}\right)^{n}\;
\MRM{H}_{2n}\left(\frac{\phi}{\sqrt{2\,\upsilon}}\right), 
\qquad
\upsilon\;=\;
\frac{\gamma}{1-\beta^{2}}. 
\label{model.6}}
We have that $D_{Z_{UV}}(R)(P_{2n})\,=\,\lambda_{2n}\,P_{2n}$, 
where $\lambda_{2n}\,=\,\alpha\beta^{2n}\,=\,L^{\sigma_{2n}}$ 
with scaling dimension $\sigma_{2n}\,=\,D+n(2-D)\,=\,3-n$.
} 
In three dimensions, $P_{0}$, $P_{2}$, and $P_{4}$ are relevant, 
$P_{6}$ is marginal, and all others are irrelevant. From the set of 
eigenfunctions, we select the relevant non-quadratic member $P_{4}$.
Its explicit form is 
\BEG{gather}{
P_{4}(\phi,\upsilon)\,=\,
\phi^{4}-6\,\upsilon\,\phi^{2}+3\,\upsilon^{2},
\qquad
\sigma_{4}\;=\;4-D\;=\;1.
\label{model.7}}
Another way of saying that $P_{4}$ is an eigenfunction of the 
tangent map with eigenvalue $\lambda_{4}$ is the following. Let 
$V(\phi,g)\,=\,g\,P_{4}(\phi,\upsilon)$. Then we have that
\BEG{gather}{
D_{Z_{UV}}(R)(V)(\psi,\delta g)\;=\;
V(\psi,g),
\qquad
\delta^{-1}\;=\; 
\alpha\beta^{4}\;=\;
L^{\sigma_{4}}.
\label{model.8}}
In other words, $V(\phi,g)$ is a fixed point of $D_{Z_{UV}}(R)
\times\delta^{\star}$, the tangent map extended by a flow of $g$. 
The linear function $g\rightarrow\delta g$ will serve as our step 
$\beta$-function. 

As in \cite{Wieczerkowski:1997}, the goal of this paper is to 
construct a non-linear analogue of \REF{model.8}, which is a
restricted homeomorphism from the tangent space at $Z_{UV}$ to an 
invariant cone in $\MCA{M}$, originating at $Z_{UV}$.

\subsection{Extended renormalization group}

From now on, we turn our attention from points in $\MCA{M}$ to 
parametrized curves in $\MCA{M}$, originating at $Z_{UV}$.
\BEG{DEF}{Let $g_{\MRM{max}}\in\MBB{R}^{+}$ and $\MBB{G}\,=\,
[0,g_{\MRM{max}}]$. 
Let $\MCA{N}$ be the space of continuous functions
$Z:\MBB{R}\times\MBB{G}\rightarrow\MBB{R}$ such that
$Z(\phi,g)\,=\,Z(-\phi,g)$ for all $(\phi,g)\in\MBB{R}\times\MBB{G}$,
$Z(\phi,0)\,=\,Z_{UV}(\phi)$, and $\| Z\|_{\infty}\,=\,
\sup_{(\phi,g)\in\MBB{R}\times\MBB{G}}\vert Z(\phi,g)\vert$ 
is finite.}
For technical reasons, we introduced a maximal coupling $g_{\MRM{max}}$. 
By default, this maximal coupling is an arbitrary large number 
in the following. Restrictions on $g_{\MRM{max}}$ will be explicitely 
stated.
\BEG{DEF}{
Let $\sigma_{4}=4-D$ and $\delta=L^{-\sigma_{4}}$. 
Let $S$ be the non-linear transformation given by
\BEG{gather}{
S(Z)(\psi,g)\;=\;
(R\times\delta^{\star})(Z)(\psi,g)\;=\;
R(Z)(\psi,\delta g).
\label{model.9}}}
\BEG{LEM}{
The transformation $S$ acts on the space $\MCA{N}$. It satisfies the 
bound $\| S(Z)\|_{\infty}\,\leq\,\| Z\|_{\infty}^{\alpha}$. 
}
This bound is an immediate consequence of $\delta\MBB{G}\subset\MBB{G}$
(since $\delta g\,<\,g$) and
\BEG{gather}{
\sup_{(\phi,g)\in\MBB{R}\times\MBB{G}}
\vert S(Z)(\phi,g)\vert\;\leq\;
\left\{
   \sup_{(\phi,g)\in\MBB{R}\times\delta\MBB{G}}
   \vert Z(\phi,g)\vert
\right\}^{\alpha}
\label{model.10}} 
Functions in $\MCA{N}$ are bounded but not necessarily decaying 
at large fields.
To bound non-perturbative contributions, we will need an exponential
decay at large fields. The following observation suggests how to select 
a subspace $\MCA{N}_{QU}\subset\MCA{N}$, which suits this purpose.
\BEG{LEM}{Let $Z_{QU}:\MBB{R}\times\MBB{G}\rightarrow
\MBB{R}^{+}$ be the positive continuous function given by
\BEG{gather}{
Z_{QU}(\phi,g)\,=\,
\exp\left\{a_{QU}(g)-\frac{b_{QU}(g)}{2}\phi^{2}\right\},
\label{model.11}}
where
\BEG{gather}{
a_{QU}(g)\,=\,
\frac{\alpha-1}{2\,\alpha}\,
\sum_{n=1}^{\infty}\alpha^{-n}\,
\log\left\{
\frac{1+\left(\delta^{-n}\,g\right)^{\rho}}
{1+g^{\rho}}
\right\}
\label{model.12}}
and
\BEG{gather}{
b_{QU}(g)\,=\,
\frac{\alpha\beta^{2}-1}{\alpha\gamma}\,
\frac{g^{\rho}}{1+g^{\rho}},
\label{model.13}}
where $\rho\,=\,\frac{2}{4-D}\,=\,2$.
The function $Z_{QU}$ is an element of $\MCA{N}$. 
It is a fixed point of $S$.}
The assignment $g\mapsto Z_{QU}(\cdot,g)$ is a continuous 
parametrized curve in $\MCA{M}^{+}$. There is no reason, not to set
$\MBB{G}\,=\,\MBB{R}^{+}$ at this point (or $g_{\MRM{max}}\,=\,\infty$). 
Then $Z_{QU}$ becomes a curve of Gauss-functions, which connects the 
two trivial fixed points of $R$, the ultraviolet 
fixed point $Z_{QU}(\phi,0)\,=\,Z_{UV}(\phi)$ to the high temperature 
fixed point $Z_{QU}(\phi,\infty)\,=\,Z_{HT}(\phi)$. 
\BEG{DEF}{Let $\MCA{N}_{QU}$ be the subspace of $\MCA{N}$, consisting 
of functions with finite norm
\BEG{gather}{
\|Z\|_{QU}\;=\;
\sup_{(\phi,g)\in\MBB{R}\times\MBB{G}}
\left\vert\frac{Z(\phi,g)}{Z_{QU}(\phi,g)}\right\vert,
\label{model.14}}
completed to a Banach space.}
Functions in $\MCA{N}_{QU}$ are in particular continuous in $g$. 
Their most useful property is the inbuilt bound
\BEG{gather}{
\vert Z(\phi,g)\vert\;\leq\;\|Z\|_{QU}\;Z_{QU}(\phi,g).
\label{model.15}}
At fixed $g$, it compares the decay of $Z$ at large $\phi$ with the 
decay of the fixed point $Z_{QU}$. 
\REF{model.15} serves as the basic large field bound in this paper.
\BEG{LEM}{The non-linear transformation $S$ acts on $\MCA{N}_{QU}$. 
(It leaves invariant $\MCA{N}_{QU}\subset\MCA{N}$.) Let 
$Z\in\MCA{N}_{QU}$. Then $\|S(Z)\|_{QU}\,\leq\,\|Z\|_{QU}^{\alpha}$.
}
We can now state our goal more precisely as to find a 
non-trivial fixed point of $S$ in $\MCA{N}_{QU}$ (besides the 
$g$-independent trivial fixed points of $R$, and besides $Z_{QU}$). 
To this aim, we need to introduce yet another subspace of 
$\MCA{N}_{QU}$.
\BEG{DEF}{Let $g_{\MRM{max}}\,<\,\infty$. Let $F:\MBB{G}\rightarrow
\MBB{R}^{+}$ be a continuous positive function with finite norm 
$\| F\|_{\infty}\,=\,\sup_{g\in\MBB{G}}\vert F(g)\vert$. Let 
$\MCA{N}_{F}$ be the subspace of $\MCA{N}_{QU}$ consisting of
functions with finite norm
\BEG{gather}{
\| Z\|_{F}\;=\;
\sup_{(\phi,g)\in\MBB{R}\times\MBB{G}}
\left\vert
\frac{Z(\phi,g)}{Z_{QU}(\phi,g)\,F(g)}
\right\vert.
\label{model.16}}
}
\REF{model.16} refines \REF{model.15}. Its motive is
to gain control over the $g$-dependence of $Z$.\footnote{The 
information that $Z$ is an element of $\MCA{N}_{QU}$ implies no more 
information about its $g$-dependence than 
that, for all $\phi\in\MBB{R}$, $Z(\phi,g)$ is a bounded 
function of $g$.} It will be crucial to carefully choose the 
function $F$. To give a flavour of, what kind of function $F$ should 
be, notice the following fact. $Z\in\MCA{N}_{F}$ implies that
\BEG{gather}{
\vert Z(\phi,g)\vert\;\leq\;
\| Z\|_{F}\;Z_{QU}(\phi,g)\;F(g).
\label{model.17}}
and thus
\BEG{gather}{
\vert S(Z)(\phi,g)\vert\;\leq\;
\| Z\|_{F}^{\alpha}\;Z_{QU}(\phi,g)\;F(\delta\,g)^{\alpha}.
\label{model.18}}
If $F$ is such that the condition $F(\delta\,g)^{\alpha}\,\leq\,\,F(g)$ 
holds for all $g\in\MBB{G}$, then $\MCA{N}_{F}$ is invariant under $S$. 
Furthermore, the ball $\|Z\|_{F}\,\leq\,1$ is then mapped to itself
under $S$. Perturbation theory suggests functions of the form 
$F(g)\,=\,C\,g^{\sigma}\,\exp\left(c\,g^{\tau}\right)$. Depending 
on the actual values of $C$, $\sigma$, and $\tau$, the above condition 
may pose a restriction on $g_{\MRM{max}}$.

\section{Contraction mapping}

In this section, we present our method to construct a non-trivial 
fixed point of $S$. As in \cite{Wieczerkowski:1997}, we split 
$Z\,=\,Z_{1}+Z_{2}$ into two parts $Z_{1}$ and $Z_{2}$, 
where $Z_{1}$ is an approximate 
fixed point, and where $Z_{2}$ is an error term. $Z_{1}$ will be 
kept fixed. We intend to estimate the non-linear mapping of the 
$Z_{2}$ under certain assumptions on $Z_{1}$. 
\BEG{LEM}{Let $Z_{1}$ and $Z_{2}$ be two elements of $\MCA{N}_{QU}$. 
Then 
\BEG{gather}{
S(Z_{1}+Z_{2})-Z_{1}\;=\;
T_{1}(Z_{1})+T_{2}(Z_{1},Z_{2}),
\label{contract.1}}
where
\BEG{gather}{
T_{1}(Z_{1})\;=\;
S(Z_{1})-Z_{1}
\label{contract.2}}
and
\BEG{gather}{
T_{2}(Z_{1},Z_{2})\;=\;
S(Z_{1}+Z_{2})-S(Z_{1}).
\label{contract.3}}}
Since $S$ acts on $\MCA{N}_{QU}$, the mappings $T_{1}$ and $T_{2}$
are both well defined. The norm of $T_{1}(Z_{1})$ measures the 
quality of the approximate fixed point $Z_{1}$. Equipped with a bound 
on $T_{1}(Z_{1})$, the issue is to find bounds on $Z_{2}$ which
iterate under \REF{contract.1}.

\subsection{Invariant ball $\MCA{B}$}

The dependence of $Z_{1}$ and $Z_{2}$ on $g$ will be controlled 
with two different functions $F_{1}$ and $F_{2}$ to be specified below.
\BEG{LEM}{Let $F_{i}:\MBB{G}\rightarrow\MBB{R}^{+}$, $i\,=\,1,2$, 
be two continuous positive functions. Abbreviate $\MCA{N}_{i}\,=\,
\MCA{N}_{F_{i}}$ and $\|Z_{i}\|\,=\,\| Z_{i}\|_{F_{i}}$. Assume
two functions $Z_{i}$ with $Z_{i}\in\MCA{N}_{i}$. Thus for all 
$(\phi,g)\in\MBB{R}\times\MBB{G}$,
\BEG{gather}{
\vert Z_{i}(\phi,g)\vert\;\leq\;
\| Z_{i}\|\;Z_{QU}(\phi,g)\;F_{i}(g).
\label{contract.4}}
Then $T_{2}(Z_{1},Z_{2})$ is an element of $\MCA{N}_{QU}$.
For all $(\phi,g)\in\MBB{R}\times\MBB{G}$ we have that
\BEG{gather}{
\vert T_{2}(Z_{1},Z_{2})(\phi,g)\vert\;\leq\;
\nonumber\\
\biggl\{
   \biggl(
      \|Z_{1}\|\,F_{1}(\delta g)
      +\|Z_{2}\|\,F_{2}(\delta g)
   \biggr)^{\alpha}
   -\biggl(
      \|Z_{1}\|\,F_{1}(\delta g)
   \biggr)^{\alpha}
\biggr\}\;
Z_{QU}(\phi,g).
\label{contract.5}} 
\label{bound.1}}
{\bf Proof}\hspace{1mm}   
$T_{2}(Z_{1},Z_{2})$ has the integral representation
\BEG{align}{
&T_{2}(Z_{1},Z_{2})(\psi,g)\;=\;
\int_{0}^{1}\MRM{d}t\;
\PAR{t}\;
S(Z_{1}+t\,Z_{2})(\psi,g)
\nonumber\\&=\;
\int_{0}^{1}\MRM{d}t\;
\PAR{t}\;
\left\{
   \int\MRM{d}\mu_{\gamma}(\zeta)\;
   \left(Z_{1}+t\,Z_{2}\right)\hspace{-2pt}(\beta\psi+\zeta,\delta g)
\right\}^{\alpha}
\nonumber\\&=
\int_{0}^{1}\MRM{d}t\;
\alpha\;
\left\{
   \int\MRM{d}\mu_{\gamma}(\zeta)\;
   \left(Z_{1}+t\,Z_{2}\right)\hspace{-2pt}(\beta\psi+\zeta,\delta g)
\right\}^{\alpha-1}\;
\nonumber\\&\phantom{=}\times
\int\MRM{d}\mu_{\gamma}(\zeta)\;
Z_{2}(\beta\psi+\zeta,\delta g)
\label{contract.6}}
Using the bound \REF{contract.4}, it follows that
\BEG{align}{
&\vert T_{2}(Z_{1},Z_{2})(\psi,g)\vert
\nonumber\\&\leq\;
\int_{0}^{1}\MRM{d}t\;
\alpha\;
\biggl(
   \|Z_{1}\|\,F_{1}(\delta g)
   +t\,\|Z_{2}\|\,\,F_{2}(\delta g)
\biggr)^{\alpha-1}\;
\nonumber\\&\phantom{\leq}\times
\|Z_{2}\|\;F_{2}(\delta g)\;
\left\{
   \int\MRM{d}\mu_{\gamma}(\zeta)\;
   Z_{QU}(\beta\psi+\zeta,\delta g)
\right\}^{\alpha}
\nonumber\\&=\;
\int_{0}^{1}\MRM{d}t\;
\PAR{t}\;
\biggl(
   \|Z_{1}\|\,F_{1}(\delta g)
   +t\,\|Z_{2}\|\,F_{2}(\delta g)
\biggr)^{\alpha}\;
Z_{QU}(\psi,g)
\nonumber\\&=\;
\biggl\{
\biggl(
   \|Z_{1}\|\,F_{1}(\delta g)
   +\|Z_{2}\|\,F_{2}(\delta g)
\biggr)^{\alpha}
-\biggl(
   \|Z_{1}\|\,F_{1}(\delta g)
\biggr)^{\alpha}
\biggr\}\;
\nonumber\\&\phantom{\leq}\times
Z_{QU}(\psi,g).\quad\Box
\label{contract.7}}
Differing from \cite{Gawedzki/Kupiainen:1984}, no split into small 
and large fields is required here. Small and large fields are taken 
care of simultaneously by the $\phi$-dependence of $Z_{QU}$. 
For the right hand side of \REF{contract.5}, we notice the elementary
estimate
\BEG{gather}{
\biggl(
   \|Z_{1}\|\,F_{1}(\delta g)
   +\|Z_{2}\|\,F_{2}(\delta g)
\biggr)^{\alpha}
-\biggl(
   \|Z_{1}\|\,F_{1}(\delta g)
\biggr)^{\alpha}\;\leq
\nonumber\\
\alpha\;
\biggl(
   \|Z_{1}\|\,F_{1}(\delta g)
   +\|Z_{2}\|\,F_{2}(\delta g)
\biggr)^{\alpha-1}\;
\|Z_{2}\|\;F_{2}(\delta g).
\label{contract.8}}
\BEG{LEM}{Let $Z_{i}$ be as in Lemma \ref{bound.1}. Let
$Z_{1}$ be such that in addition 
$\|T_{1}(Z_{1})\|\,=\,\|T_{1}(Z_{1})\|_{F_{2}}$
is finite.
Then we have that for all $(\phi,g)\in\MBB{R}\times\MBB{G}$, 
\BEG{gather}{
\vert T_{1}(Z_{1})(\phi,g)\vert\;\leq\;
\| T_{1}(Z_{1})\|\;
Z_{QU}(\phi,g)\;F_{2}(g).
\label{contract.9}}
Let $n$ be an integer, $n\in\{1,2,3,\ldots\}$. 
Assume that $F_{1}$ and $F_{2}$ conspire such that 
\BEG{gather}{
\alpha\;
\biggl(
   \|Z_{1}\|\,F_{1}(\delta g)
   +(n+1)\;\|T_{1}(Z_{1})\|\,F_{2}(\delta g)
\biggr)^{\alpha-1}\;
\,F_{2}(\delta g)\;\leq\;
\frac{n}{n+1}\;F_{2}(g).
\label{contract.10}}
Let $\MCA{B}$ be the ball in $\MCA{N}_{2}$ given by
\BEG{gather}{
\| Z_{2}\|\;\leq\;(n+1)\;\|T_{1}(Z_{1})\|.
\label{contract.11}}
Then it follows that $\MCA{B}$ is invariant under the 
transformation \REF{contract.1}. For all $Z_{2}\in\MBB{B}$ 
and for all $(\phi,g)\in\MBB{R}\times\MBB{G}$,
\BEG{gather}{
\left\vert
S(Z_{1}+Z_{2})(\phi,g)-Z_{1}(\phi,g)
\right\vert\;\leq\;
(n+1)\;\|T_{1}(Z_{1})\|\;
Z_{QU}(\phi,g)\;F_{2}(\phi,g).
\label{contract.12}}
\label{bound.2}}
{\bf Proof}\hspace{1mm}   
From \REF{contract.5}, \REF{contract.8}, \REF{contract.10}, and
\REF{contract.11}, it follows that 
\BEG{align}{
\|S(Z_{1}+Z_{2})-Z_{1}\|&\leq\;
\|T_{1}(Z_{1})\|
+\|T_{2}(Z_{1},Z_{2})\|
\nonumber\\&\leq\;
\|T_{1}(Z_{1})\|
+\frac{n}{n+1}\;
\|Z_{2}\|
\nonumber\\&\leq\;
\|T_{1}(Z_{1})\|
+\frac{n}{n+1}\;
(n+1)\;\|T_{1}(Z_{1})\|
\nonumber\\&=\;
(n+1)\;\|T_{1}(Z_{1})\|.
\quad\Box
\label{contract.13}}
We remark that the additional assumptions made in Lemma \ref{bound.2} 
do not clash with those made in Lemma \ref{bound.1}.

Lemma \ref{bound.2} suggests the following strategy. We look for 
functions $Z_{1}$, $F_{1}$, and $F_{2}$ such that 
$\|Z_{1}\|_{F_{1}}$ and $\|T_{1}(Z_{1})\|_{F_{2}}$ are both 
finite. Furthermore, $F_{1}$ and $F_{2}$ have to satisfy 
\REF{contract.10}. Then we have 
an invariant ball of error terms $Z_{2}$ in the norm associated
with $F_{2}$.

\subsection{Contraction mapping}

Our next task is to establish the contraction property for
the transformation \REF{contract.1} on the ball $\MCA{B}$.  
\BEG{LEM}{Let $Z_{1}$, $Z_{2}$, and $Z_{2}^{\prime}$ be as in 
Lemma \ref{bound.1}. Then we have for all $(\phi,g)\in\MBB{R}\times
\MCA{G}$ the bound
\BEG{gather}{
\left\vert
   S(Z_{1}+Z_{2}^{\prime})(\phi,g)
   -S(Z_{1}+Z_{2})(\phi,g)
\right\vert\;\leq
\nonumber\\
\alpha\;
\biggl(
   \|Z_{1}\|\,F_{1}(\delta g)+
   \sup_{t\in [0,1]}
   \left\|
      Z_{2}+t\,(Z_{2}^{\prime}-Z_{2})
   \right\|\,F_{2}(\delta g)
\biggr)^{\alpha-1}
F_{2}(\delta g)\;
\nonumber\\ \times\;
Z_{QU}(\psi,g)\;
\|Z_{2}^{\prime}-Z_{2}\|
\label{contract.14}}
}
{\bf Proof}\hspace{1mm}   
From the integral representation
\BEG{align}{
&S(Z_{1}+Z_{2}^{\prime})(\psi,g)-S(Z_{1}+Z_{2})(\psi,g)
\nonumber\\&=
\int_{0}^{1}\MRM{d}t\;\PAR{t}\;
S\bigl(
   Z_{1}+Z_{2}+t\,(Z_{2}^{\prime}-Z_{2})
\bigr)(\psi,g)
\nonumber\\&=
\int_{0}^{1}\MRM{d}t\;\PAR{t}\;
\left\{
   \int\MRM{d}\mu_{\gamma}(\zeta)\;
   \bigl(
      Z_{1}+Z_{2}+t\,(Z_{2}^{\prime}-Z_{2})
   \bigr)(\beta\psi+\zeta,\delta g)
\right\}^{\alpha}
\nonumber\\&=
\int_{0}^{1}\MRM{d}t\;
\alpha\;
\left\{
   \int\MRM{d}\mu_{\gamma}(\zeta)\;
   \bigl(
      Z_{1}+Z_{2}+t\,(Z_{2}^{\prime}-Z_{2})
   \bigr)(\beta\psi+\zeta,\delta g)
\right\}^{\alpha-1}
\nonumber\\&\phantom{=}\times
\int\MRM{d}\mu_{\gamma}(\zeta)\;
   \bigl(
      Z_{2}^{\prime}-Z_{2}
   \bigr)(\beta\psi+\zeta,\delta g)
\label{contract.15}}
it follows that
\BEG{align}{
&\left\vert
   S(Z_{1}+Z_{2}^{\prime})(\psi,g)-S(Z_{1}+Z_{2})(\psi,g)
\right\vert
\nonumber\\&\leq\;
\int_{0}^{1}\MRM{d}t\;
\alpha\;
\biggl(
   \|Z_{1}\|\,F_{1}(\delta g)+
   \left\|
      Z_{2}+t\,(Z_{2}^{\prime}-Z_{2})
   \right\|\,F_{2}(\delta g)
\biggr)^{\alpha-1}
\nonumber\\&\phantom{=}\times
\;F_{2}(\delta g)\;
\left\{
   \int\MRM{d}\mu_{\gamma}(\zeta)\;Z_{QU}(\beta\psi+\zeta,\delta g)
\right\}^{\alpha}
\;\|Z_{2}^{\prime}-Z_{2}\|
\nonumber\\&\leq
\alpha\,
\biggl(
   \|Z_{1}\|\,F_{1}(\delta g)+
   \sup_{t\in [0,1]}
   \left\|
      Z_{2}+t\,(Z_{2}^{\prime}-Z_{2})
   \right\|\,F_{2}(\delta g)
\biggr)^{\alpha-1}
F_{2}(\delta g)\;
\nonumber\\&\phantom{=}\times
\;Z_{QU}(\psi,g)\;\;
\|Z_{2}^{\prime}-Z_{2}\|.
\quad\Box
\label{contract.16}}
The contraction property follows from this estimate in 
cooperation with Lemma \ref{bound.2}.
\BEG{LEM}{
Let $Z_{1}$, $Z_{2}$, and $Z_{2}^{\prime}$ be as in Lemma
\ref{bound.2}. For all $Z_{2}\in\MCA{B}$ and $Z_{2}^{\prime}\in
\MCA{B}$, we have that
\BEG{gather}{
\|S(Z_{1}+Z_{2}^{\prime})-S(Z_{1}+Z_{2})\|\;\leq\;
\frac{n}{n+1}\;\|Z_{2}^{\prime}-Z_{2}\|.
\label{contract.17}}
}
{\bf Proof}\hspace{1mm}   
Since $\MCA{B}$ is convex, it follows that for all $t\in[0,1]$, 
\BEG{gather}{
\left\|
   Z_{2}+t\,(Z_{2}^{\prime}-Z_{2})
\right\|\,F_{2}(\delta\,g)\;\leq\;
(n+1)\;
\|T_{1}(Z_{1})\|.
\label{contract.18}}
We can therefore use \REF{contract.11} to conclude the asserted
bound from \REF{contract.14}, namely
\BEG{gather}{
\left\vert
   S(Z_{1}+Z_{2}^{\prime})(\phi,g)
   -S(Z_{1}+Z_{2})(\phi,g)
\right\vert\;
\nonumber\\
\leq
\frac{n}{n+1}\;Z_{QU}(\phi,g)\;F_{2}(g)\;
\|Z_{2}^{\prime}-Z_{2}\|.
\quad\Box
\label{contract.19}}
\BEG{COR}{Let $\MCA{B}$ be as in Lemma \ref{bound.2}. $\MCA{B}$ is
invariant under the transformation \REF{contract.1}. 
The transformation \REF{contract.1} is a contraction mapping.
The transformation \REF{contract.1} has a unique fixed point in 
$\MCA{B}$. 
}
The iteration of the contraction mapping is a convergent 
scheme to compute the fixed point.
Let $Z_{2,0}\,=\,0$ and define a sequence of (non-perturbative) 
approximants by
\BEG{gather}{
Z_{2,n+1}\;=\;S(Z_{1}+Z_{2,n})-Z_{1}.
\label{contract.20}} 
As $n\rightarrow\infty$, this sequence converges to the desired 
fixed point. We cannot say much about this limit at this level of 
generality. However, if $Z_1$ is strictly positive than it follows 
immediately that the fixed point is non-negative. 

\subsection{Estimates for all couplings}

Our next subject is to construct examples of the two functions 
$F_{1}$ and $F_{2}$ with the property \REF{contract.10}. 
These examples are tailor made for perturbation theory. 

Suppose that we are told an approximate fixed point $Z_{1}$ from 
an independent calculation, whose large field decay is (at least) 
Gaussian. Then we proceed as follows. 
\begin{enumerate}
\item
We determine $F_{1}$ such that
\BEG{gather}{
\sup_{\phi\in\MBB{R}}
\left\vert
   \frac{Z_{1}(\phi,g)}{Z_{QU}(\phi,g)}
\right\vert\;\leq\;
F_{1}(g).
\label{all.1}}
Then $\| Z_{1}\|_{F_{1}}\,\leq\,1$. Since it contains 
information about the large field behavior of $Z_{1}$, 
we call this bound the {\sl stability bound} on $Z_{1}$. 
\item
We determine $F_{2}$ from an estimate on $T_{1}(Z_{1})$ 
such that 
\BEG{gather}{
\sup_{\phi\in\MBB{R}}
\left\vert
   \frac{T_{1}(Z_{1})(\phi,g)}{Z_{QU}(\phi,g)}
\right\vert\;\leq\;
C_{1}\;F_{2}(g)
\label{all.2}}
for some finite constant $C_{1}$. 
Then $\|T_{1}(Z_{1})\|_{F_{2}}\,\leq\,C_{1}$.
We call this second bound 
the {\sl error bound} on $Z_{1}$.
\item 
We check that $F_{1}$ and $F_{2}$ satisfy \REF{contract.10}.
\end{enumerate}
If $F_{1}$ and $F_{2}$ are functions of the type of this 
section, then the validity of \REF{contract.10} is 
guaranteed by the below estimates.

Remember that $Z_{1}$ plays the role of an approximate fixed point.
This means that $T_{1}(Z_{1})$ should be small compared with $Z_{1}$.
A construction for all couplings should in particular cover the 
case of small couplings. Inspired by perturbation theory, we choose 
\BEG{gather}{
F_{2}(g)\;=\;
g^{\sigma}\;F_{1}(g).
\label{all.3}}
It says that the error term goes to zero as the coupling goes to 
zero. The speed of this process is given by the exponent $\sigma$. 
\BEG{LEM}{Let $F_{1}:\MBB{G}\rightarrow\MBB{R^{+}}$ be a 
positive continuous function. Let $\|Z_{1}\|_{F_{1}}\leq 1$. Let 
$\sigma$ be a positve real number, with $\sigma\,>\sigma_{\star}$,
where\footnote{The exponent $\sigma_{\star}$ is such that
$\alpha\delta^{\sigma_{\star}}=1$. For $\sigma>\sigma_{\star}$
the block volume $\alpha$ is beaten by the power of the 
step $\beta$-function.} 
\BEG{gather}{
\sigma_{\star}\;=\;
\frac{D}{4-D}\;=\;3.
\label{all.4}}
Let $F_{2}(g)=g^{\sigma}\,F_{1}(g)$. Let 
$\|T_{1}(Z_{1})\|_{F_{2}}\leq C_{1}$ for some positive constant
$C_{1}$. Let $n\in\MBB{N}$. Put $C_{2}=(n+1)C_{1}$.
Assume that $F_{1}$ has the following property. Let there exists a 
positive constant $C_{F}$ such that for all $g\in\MBB{G}$, 
\BEG{gather}{
\MRM{e}^{(\alpha-1)\,C_{2}\,(\delta g)^{\sigma}}\;
F_{1}(\delta g)^{\alpha}\;\leq\;
C_{F}\;F_{1}(g).
\label{all.5}} 
Then we have the following estimate. 
There exists a positive real number $L_{\MRM{min}}$ such that for $L$ 
larger than this number, we have the bound
\BEG{gather}{
\alpha\;
\biggl(
   \|Z_{1}\|\,F_{1}(\delta\,g)
   +(n+1)\,\|T_{1}(Z_{1})\|\,F_{2}(\delta\,g)
\biggr)^{\alpha-1}\;
F_{2}(\delta\,g)\;\leq\;
\frac{n}{n+1}\;F_{2}(g).
\label{all.6}}
}
{\bf Proof}\hspace{1mm}   
From the assumptions
\REF{all.2}, \REF{all.3}, and \REF{all.5}, it follows
that
\BEG{align}{
&\alpha\;
\biggl(
   \|Z_{1}\|\,F_{1}(\delta g)
   +(n+1)\,\|T_{1}(Z_{1})\|\,F_{2}(\delta g)
\biggr)^{\alpha-1}\;
F_{2}(\delta g)
\nonumber\\&\leq\;
\alpha\;
\biggl(
1+C_{2}\,(\delta g)^{\sigma}
\biggr)^{\alpha-1}\;
(\delta g)^{\sigma}\;
F_{1}(\delta g)^{\alpha}
\nonumber\\[1mm]&\leq\;
\alpha\;
\MRM{e}^{(\alpha-1)\,C_{2}\,(\delta g)^{\sigma}}\;
(\delta g)^{\sigma}\;
F_{1}(\delta g)^{\alpha}
\nonumber\\[2mm]&\leq\;
\alpha\;(\delta g)^{\sigma}\;C_{F}\;F_{1}(g)
\nonumber\\[2mm]&=\;
\alpha\delta^{\sigma}\;C_{F}\;F_{2}(g).
\label{all.7}}
For $\sigma\,>\,\sigma_{\star}$, the $L$-dependent factor 
$\alpha\delta^{\sigma}\;=\;L^{D+(D-4)\,\sigma}$ can be made
arbitrary small by taking $L$ to be large. $\Box$\\[2mm]
We learn that \REF{contract.10} (identical with \REF{all.6})
holds provided that $F_{1}$ satisfies \REF{all.5}.  
An example of a function $F_{1}$ which satisfies \REF{all.5}
without a restriction on $g_{\MRM{max}}$ is the following.
\BEG{LEM}{Let $c_{\star}$ and $c$ be positive constants. 
Let $\sigma$ be a positive constant such that $\sigma\,>\,
\sigma_{\star}$. Let $F_{1}$ be the function
\BEG{gather}{
F_{1}(g)\;=\;
\exp\left(
   c_{\star}\,g^{\sigma_{\star}}
   +c\,g^{\sigma}
\right).
\label{all.8}} 
For all positive constants $C_{2}$ such that 
\BEG{gather}{
C_{2}\;\leq\;
\frac{1-\alpha\delta^{\sigma}}{\alpha\delta^{\sigma}}\;c,
\label{all.9}}
this function $F_{1}$ satisfies the bound
\BEG{gather}{
\MRM{e}^{(\alpha-1)\,C_{2}\,(\delta g)^{\sigma}}\;
F_{1}(\delta\,g)^{\alpha}\;\leq\;
F_{1}(g).
\label{all.10}}
\label{restrict}}
{\bf Proof}\hspace{1mm}   
For this particular function $F_{1}$, we have that
\BEG{align}{
&\exp\left\{
   (\alpha-1)\,C_{2}\,(\delta g)^{\sigma}
\right\}\;
\exp\left\{
   c_{\star}\,g^{\sigma_{\star}}
   +\alpha\,c\,(\delta\,g)^{\sigma}
\right\}
\nonumber\\[1mm]&\leq\;
\exp\left\{
   c_{\star}\,g^{\sigma_{\star}}
   +\alpha\delta^{\sigma}(C_{2}+c)\,g^{\sigma}
\right\}
\nonumber\\[1mm]&\leq\;
\exp\left\{
   c_{\star}\,g^{\sigma_{\star}}
   +c\,g^{\sigma}
\right\}.
\quad\Box
\label{all.11}}
Lemma \ref{restrict} contains a restriction on $C_{2}$. To avoid a 
conflict with \REF{contract.10}, the constant $c$ has to
be chosen to be sufficiently large. 

The pair $F_{i}$ given by \REF{all.8} and \REF{all.3} (with
$\sigma>\sigma_{\star}$) satisfies all properties needed for the 
contraction mapping. Remarkably, it involves no restriction on the 
size of $g_{\MRM{max}}$. With this pair, we could construct the
$\phi^{4}$-theory at arbitrary large couplings. The problem is 
to compute an approximate fixed point $Z_{1}$,
which satisfies the stability bound \REF{all.1} with this 
function $F_{1}$, and which satisfies the error bound with a
sufficiently large exponent $\sigma$. Unfortunately, we have not 
succeeded to prove these bounds for approximants from perturbation 
theory.

\subsection{Estimates for small couplings}

Therefore, we supplement the bounds for all couplings by 
suitable bounds for small couplings. There are two sources 
of constraints on the value of $g_{\MRM{max}}$. One source 
is that we may not be able to prove the stability bound on 
$Z_{1}$ for arbitrary large couplings. Also we may not be 
able to prove the error bound on $T_{1}(Z_{1})$ for 
arbitrary large couplings. In this section, we will not speak
about this source of problems. These constraints cannot be 
addressed before we actually compute $Z_{1}$. The second 
source is that the function $F_{1}$, which we find from 
\REF{all.1} (say by defining $F_{1}$ by equality), might not
meet the requirement \REF{all.5} for arbitrary large 
couplings. If both effects come together, we have to put
$g_{\MRM{max}}$ equal to the minimum from these constraints.
\BEG{LEM}{Let $c_{\star}$ and $c$ be a positive constants. Let 
$\tau$ be a positive constant such that $\tau\,<\,\sigma_{\star}$. 
Let $F_{1}$ be the function
\BEG{gather}{
F_{1}(g)\;=\;
\exp\left(c_{\star}\,g^{\sigma_{\star}}+c\,g^{\tau}\right).
\label{small.1}}
Let $C_{2}$ and $C_{F}$ be positive constants, with $C_{F}\,>\,1$.
For all values of $L$, there exists a maximal coupling $g(L)$ 
such that, for all $g$ less than this maximal coupling, we have 
that
\BEG{gather}{
\MRM{e}^{(\alpha-1)\,C_{2}\,(\delta g)^{\sigma}}\;
F_{1}(\delta g)^{\alpha}\;\leq\;
C_{F}\,F_{1}(g).
\label{small.2}}
} 
{\bf Proof}\hspace{1mm}   
The estimate is very similar to the one for all couplings. The 
identity
\BEG{gather}{
\MRM{e}^{(\alpha-1)\,C_{2}\,(\delta g)^{\sigma}}\;
F_{1}(\delta g)^{\alpha}\;=\;
\MRM{e}^{c\,(\alpha\delta^{\tau}-1)\,g^{\tau}}\;
F_{1}(g)
\label{small.3}}
yields the condition
\BEG{gather}{
\MRM{e}^{(\alpha-1)\,C_{2}\,(\delta g)^{\sigma}
+c\,(\alpha\delta^{\tau}-1)\,g^{\tau}}\;\leq\;
C_{F}
\label{small.4}}
on $g$. Given exponents $\sigma$ and $\tau$, together with 
constants $C_{2}$ and $c$, and a scale $L$, \REF{small.4}
determines $g(L)$. $\Box$\\[2mm]
To get an idea how $g(L)$ behaves as a function of $L$, 
we neglect the the first term in \REF{small.4}. Then 
\BEG{gather}{
g(L)\;\leq\;
\left(
   \frac{\ln \left(C_{F}\right)}
   {c\,(\alpha\delta^{\tau}-1)}
\right)^{\frac{1}{\tau}}
\label{small.5}}
shows that $g(L)$ shrinks as a certain power of $L$.
A typical value of $\tau$ in our perturbation theory is one.
Since $\sigma_{\star}\,=\,3$ in three dimensions, we are in
the small coupling case. The good news is that the maximal 
coupling is not necessarily ridiculously small. 

It is likely that there exist other functions $F_{i}$ 
which meet the requirements of the above contraction 
mapping. The particular ones considered here are made for
approximants $Z_{1}$ from perturbation theory. The interesting
question remains whether this contraction mapping works with 
other approximants, for instance approximants from numerical
work. This would be more in the spirit of the fixed point
construction of \cite{Koch/Wittwer:1986}.

\section{Linear approximation}

The simplest approximation is a pure $\phi^{4}$-vertex.
In this section, we will prove a stability bound and an error 
bound for this linear approximation defined by
\BEG{gather}{ 
Z(\phi,g)\;=\;
\MRM{e}^{-V(\phi,g)},
\qquad
V(\phi,g)\;=\;g\,P_{4}(\phi,\upsilon).
\label{lin.1}}
(To simplify the notation, we write $Z$ instead of $Z_{1}$.)
The linear approximation will not suffice to obtain a 
contraction mapping in three dimensions. But it is a part of, 
and also an instructive example for, the finer bounds to be 
presented below.

\subsection{Stability bound}

The classical stability bound for \REF{lin.1} relies on 
analyticity in the field variable a strip around the real 
axis \cite{Gawedzki/Kupiainen:1984}. We proceed differently 
therefrom \cite{Pordt:1990,Wieczerkowski:1997}. 
\BEG{LEM}{The polynomial $P_{4}(\phi,\upsilon)$ is bounded 
from below by 
\BEG{gather}{
P_{4}(\phi,\upsilon)\;\geq\;
\frac{1}{2}\,\phi^{4}-15\,\upsilon^{2}.
\nonumber\\[-1mm]
\label{lin.2}}
\label{stab.1}}
{\bf Proof}\hspace{1mm}   
\BEG{align}{
P_{4}(\phi,\upsilon)&=\;
\phi^{4}-6\,\upsilon\,\phi^{2}+3\,\upsilon^{2}
\nonumber\\&=\;
\left(\epsilon\,\phi^{2}-\frac{3\,\upsilon}{\epsilon}\right)^{2}
-\left(\frac{3\,\upsilon}{\epsilon}\right)^{2}
+(1-\epsilon^{2})\,\phi^{4}
+3\,\upsilon^{2}
\nonumber\\&\geq\;
(1-\epsilon^{2})\,\phi^{4}+
3\,\left(1-\frac{3}{\epsilon^{2}}\right)\,\upsilon^{2}.
\label{lin.3}}
For $\epsilon^{2}\,=\,\frac{1}{2}$, the assertion follows. 
$\Box$\\[2mm]
\BEG{LEM}{For all $(\phi,g)\in\MBB{R}\times\MBB{G}$, we have that
\REF{lin.1} is bounded from above by
\BEG{gather}{
Z(\phi,g)\;\leq\;
Z_{QU}(\phi,g)\;
\MRM{e}^{a_{1}(g)}
\label{lin.4}}
with
\BEG{gather}{
a_{1}(g)\;\leq\;
15\,\upsilon^{2}\,g+\frac{1}{8}\,g^{2\rho-1}-a_{QU}(g).
\nonumber\\[-1mm]
\label{lin.5}}
\label{stab.2}}
{\bf Proof}\hspace{1mm}   
For any $c$, we have the elementary bound
\BEG{gather}{
\phi^{4}\;=\;
(\phi^{2}-\frac{c}{2})^{2}+c\,\phi^{2}-\frac{c^{2}}{4}\;\geq\;
c\,\phi^{2}-\frac{c^{2}}{4}.
\label{lin.6}}
Put $c\,=\,g^{\rho-1}$ to obtain
\BEG{gather}{
\frac{g}{2}\phi^{4}\;\geq\;
\frac{g^{\rho}}{2}\phi^2-\frac{g^{2\rho-1}}{8}.
\label{lin.7}}
The values of $L$, $D$, and $\gamma$ are such that \REF{model.13}
is bounded from above by
\BEG{gather}{
b_{QU}(g)\;=\;
\frac{L^{2}-1}{L^{D}\,\gamma}\frac{g^{\rho}}{1+g^{\rho}}\;<\;
\frac{L^{2-D}}{\gamma}\;
g^{\rho}\;<\;g^{\rho}.\quad
\Box
\label{lin.8}}
The bound \REF{lin.5} suggests a function $F_{1}$ of the form 
\REF{all.8}.
\BEG{LEM}{Let $Z$ be given by \REF{lin.1}. Let 
$F_{1}:\MBB{G}\rightarrow\MBB{R}^{+}$ be the function
\BEG{gather}{
F_{1}(g)\;=\;
\exp\left(
   15\,\upsilon^{2}\,g+\frac{g^{2\rho-1}}{8}
\right).
\label{lin.9}}
Then $Z$ is bounded in the norm \REF{model.16}. We have 
that $\|Z\|_{F_{1}}\;\leq\;1$.
\label{stab.3}}
{\bf Proof}\hspace{1mm}   
For all $g\,\geq\, 0$,  $a_{QU}(g)\,\geq\,0$, wherefore
\BEG{gather}{
\|Z\|_{F_{1}}\;=\;
\sup_{(\phi,g)\in\MBB{R}\times\MBB{G}}
\left\vert
   \frac{Z(\phi,g)}{Z_{QU}(\phi,g)\,F_{1}(g)}
\right\vert
\;\leq\;
\sup_{g\in\MBB{G}}\;
\MRM{e}^{-a_{QU}(g)}
\;\leq\;1.\quad\Box
\label{lin.10}}
In three dimensions, the value of $\rho$ is two and that of 
$\sigma_{\star}$ is three. (Recall \REF{model.13} and 
\REF{all.4}.) Coincidentally, $2\rho-1\,=\,\sigma_{\star}$.
But the second term in the exponent in \REF{lin.9} is only 
linear in $g$. Although the stability bound \REF{lin.9} holds 
for any value of $g$, it restricts our contraction mapping to 
small couplings. 

\subsection{Error bound}

Equipped with this stability bound, one is led to estimate
$T_{1}(Z)$ in the norm given by \REF{all.3}, where 
$\sigma$ is a suitable exponent. As we will see, in fact as
we know from \cite{Wieczerkowski:1997}, such a bound indeed
holds. But the exponent $\sigma$ is only one half and 
therefore smaller than $\sigma_{\star}$. The linear 
approximation therefore suffices only for a construction in
low dimensions. Let us nevertheless see how the exponent 
one half comes about. For this purpose, we 
consider the following interpolation formula.
\BEG{DEF}{Let $X:\MBB{R}\times\MBB{G}\times [0,1]\rightarrow
\MBB{R}$ be defined by
\BEG{gather}{
X(\psi,g,t)\;=\;
\Biggl\{
   \int\MRM{d}\mu_{t\,\gamma}(\zeta)\;
   \exp\biggl(
      -\int\MRM{d}\mu_{(1-t)\,\gamma}(\xi)\;
      V(\beta\psi+\zeta+\xi,\delta g)
   \biggr)
\Biggr\}^{\alpha}
\nonumber\\[-1mm]
\label{lin.11}}
}
In the limit of a vanishing covariance, the Gaussian measure 
$\MRM{d}\mu_{\gamma}(\zeta)$ becomes a Dirac measure 
$\MRM{d}\zeta\,\delta(\zeta)$. Therefore, \REF{lin.10} 
interpolates between the exponentiated linerarized 
renormalization group transformation \REF{model.5}
\BEG{gather}{
X(\psi,g,0)\;=\;
\exp\biggl(
   -\alpha\,
   \int\MRM{d}\mu_{\gamma}(\xi)\;
   V(\beta\psi+\xi,\delta g)
\biggr)
\label{lin.12}} 
and the full renormalization group transformation
\BEG{gather}{
X(\psi,g,1)\;=\;
\Biggl\{
   \int\MRM{d}\mu_{\gamma}(\zeta)\;
   \exp\biggl(
      -V(\beta\psi+\zeta,\delta g)
   \biggr)
\Biggr\}^{\alpha},
\label{lin.13}}
both transformations being extended by the flow of $g$. The 
usefulness of this interpolation relies on the following 
property.
\BEG{LEM}{For $t\in (0,1)$, $X$ is continuously differentiable 
in $t$. We have that
\BEG{gather}{
\PAR{t}X(\psi,g,t)\;=\;
\nonumber\\
\alpha\;
\Biggl\{
   \int\MRM{d}\mu_{t\,\gamma}(\zeta)\;
   \exp\biggl(
      -\int\MRM{d}\mu_{(1-t)\,\gamma}(\xi)\;
      V(\beta\psi+\zeta+\xi,\delta g)
   \biggr)
\Biggr\}^{\alpha-1}
\nonumber\\ \times
\Biggl\{
\int\MRM{d}\mu_{t\,\gamma}(\zeta)\;
\exp\biggl(
   -\int\MRM{d}\mu_{(1-t)\,\gamma}(\xi)\;
   V(\beta\psi+\zeta+\xi,\delta g)
\biggr)
\nonumber\\ \times
\frac{\gamma}{2}
\biggl(
   \PAR{\zeta}
   \int\MRM{d}\mu_{(1-t)\,\gamma}(\xi)\;
      V(\beta\psi+\zeta+\xi,\delta g)
\biggr)^{2}
\Biggr\}.
\label{lin.14}}
\label{err.1}}
{\bf Proof}\hspace{1mm}
\BEG{align}{
&\PAR{t}
\int\MRM{d}\mu_{t\,\gamma}(\zeta)\;
\exp\biggl(
   -\int\MRM{d}\mu_{(1-t)\,\gamma}(\xi)\;
   V(\phi+\zeta+\xi)
\biggr)
\nonumber\\&=\;
\int\MRM{d}\mu_{t\,\gamma}(\zeta)\;
\left[
   \frac{\gamma}{2}\frac{\partial^{2}}{\partial \zeta^{2}}
   +\PAR{t}
\right]
\exp\biggl(
   -\int\MRM{d}\mu_{(1-t)\,\gamma}(\xi)\;
   V(\phi+\zeta+\xi)
\biggr)
\nonumber\\&=\;
\int\MRM{d}\mu_{t\,\gamma}(\zeta)\;
\exp\biggl(
   -\int\MRM{d}\mu_{(1-t)\,\gamma}(\xi)\;
   V(\phi+\zeta+\xi)
\biggr)
\nonumber\\&\phantom{=}\;\times
\Biggl\{
\frac{\gamma}{2}
\biggl(
   \PAR{\zeta}
   \int\MRM{d}\mu_{(1-t)\,\gamma}(\xi)\;
      V(\phi+\zeta+\xi)
\biggr)^{2}   
\nonumber\\&\phantom{=\times}\;
-\left[
   \frac{\gamma}{2}\frac{\partial^{2}}{\partial \zeta^{2}}
   +\PAR{t}
\right]
\int\MRM{d}\mu_{(1-t)\,\gamma}(\xi)\;V(\phi+\zeta+\xi)
\Biggr\}
\label{lin.15}}
and
\BEG{gather}{
\left[
   \frac{\gamma}{2}\frac{\partial^{2}}{\partial \zeta^{2}}
   +\PAR{t}
\right]
\int\MRM{d}\mu_{(1-t)\,\gamma}(\xi)\;V(\phi+\zeta+\xi)
\;=\;0.
\quad\Box
\label{lin.16}}
Notice that \REF{lin.14} is of second order in $V$ due to 
the cancellation \REF{lin.16}. Notice furthermore that 
\REF{lin.14} is non-negative. The integral of \REF{lin.14}
yields a representation for $T_{1}(Z)$, which can be used 
to derive an upper bound of the desired form.
\BEG{LEM}{Let $Z$ be given by \REF{lin.1}. Let $X$ be 
given by \REF{lin.11}. Then we have that
\BEG{gather}{
T_{1}(Z)(\psi,g)\;=\;
X(\psi,g,1)-X(\psi,g,0)\;=\;
\int_{0}^{1}\MRM{d}t\;\PAR{t}\;X(\psi,g,t).
\label{lin.17}}
\label{err.2}}
{\bf Proof}\hspace{1mm}
Because $V$ is an eigenvector of the extended linearized 
renormalization group \REF{model.8},
\BEG{gather}{
\alpha\;
\int\MRM{d}\mu_{\gamma}(\xi)\;V(\beta\psi+\zeta,\delta\,g)
\;=\;
V(\psi,g)
\label{lin.18}}
so that $X(\psi,g,0)\,=\,Z(\psi,g)$. $\Box$\\[2mm]
A cost of this representation is that we have to repeat the 
stability analysis for the interpolated interaction. In the 
linear approximation, this follows from an explicit calculation.
The following bound is uniform in the interpolation parameter.
\BEG{LEM}{
For all $(\phi,g,t)\in\MBB{R}\times\MBB{G}\times [0,1]$, 
we have that
\BEG{gather}{
\int\MRM{d}\mu_{(1-t)\,\gamma}(\xi)\;
V(\phi+\xi,g)\;\geq\;
\frac{g}{2}\,\phi^{4}
-15\,\upsilon^{2}\,g.
\label{lin.19}}
\label{err.3}}
{\bf Proof}\hspace{1mm}
\BEG{align}{
\int\MRM{d}\mu_{(1-t)\,\gamma}(\xi)\;
V(\phi+\xi,g)&=\;
g\;
\int\MRM{d}\mu_{(1-t)\,\gamma}(\xi)\;
P_{4}(\phi+\xi,\upsilon)
\nonumber\\&=\;
g\;
P_{4}\bigl(\phi,\upsilon-(1-t)\,\gamma\bigr)
\nonumber\\&\geq\;
g\;
\left\{
   \frac{\phi^{2}}{2}
   -15\,\biggl(\upsilon-(1-t)\,\gamma\biggr)^{2}
\right\}
\label{lin.20}}
and 
\BEG{gather}{
\frac{\beta^{2}\,\gamma}{1-\beta^{2}}\;=\;
\upsilon-\gamma\;\leq\;
\upsilon-(1-t)\,\gamma\;\leq\;
\upsilon\;=\;
\frac{\gamma}{1-\beta^{2}}
\label{lin.21}}
show the assertion, since $\beta^{2}\,=\,L^{2-D}\,<\,1$.  
$\Box$

\subsubsection{Large field domination}

We are now in the position to estimate the downstairs factor 
in \REF{lin.14}, using up a fraction, say one half, of the large 
field behavior of \REF{lin.19}.
\BEG{LEM}{Let $V$ be given by \REF{lin.1}. For all 
$(\phi,g)\in\MBB{R}\times\MBB{G}$, we have that
\BEG{gather}{
\MRM{e}^{
   -\int\MRM{d}\mu_{(1-t)\,\gamma}(\xi)\;
   V(\phi+\xi,g)
}\;
\frac{\gamma}{2}
\left(
   \PAR{\phi}
   \int\MRM{d}\mu_{(1-t)\,\gamma}(\xi)\;
   V(\phi+\xi,g)
\right)^{2}\;\leq\;
\nonumber\\
C(g)\,\sqrt{g}\;
\exp\biggl(
   -\frac{g}{4}\,\phi^{4}
   +15\,g\,\upsilon^{2}
\biggr)
\label{lin.22}}
with 
\BEG{gather}{
C(g)\,\leq\,8\,\gamma\,
\biggl(
   A_{6}+9\,\upsilon^{2}\,A_{2}\,g
\biggr)
\label{lin.23}}
where
\BEG{gather}{
A_{2n}\;=\;
\sup_{\phi\in\MBB{R}}
\left(
   \MRM{e}^{-\frac{\phi^{4}}{4}}\;\phi^{2n}
\right).
\label{lin.24}}
\label{err.4}}
{\bf Proof}\hspace{1mm}
The stability bound \REF{lin.19}, in conjunction with the 
elementary estimates
\BEG{gather}{
\left(
   \PAR{\phi}P_{4}(\phi,\upsilon)
\right)^{2}\;=\;
16\;\left(
   \phi^{6}-6\,\upsilon\,\phi^{4}+9\,\upsilon^{2}\,\phi^{2}
\right)\;\leq\;
16\;\left(
   \phi^{6}+9\,\upsilon^{2}\,\phi^{2}
\right)
\nonumber\\[-1mm]
\label{lin.25}}
and
\BEG{gather}{
\exp\left(
   -\frac{g}{4}\,\phi^{4}
\right)\;
\phi^{2n}\;=\;
\exp\left\{
   -\frac{1}{4}
   \left(
      g^{\frac{1}{4}}\,\phi
   \right)^{4}
\right\}\;
\left(
   g^{\frac{1}{4}}\,\phi
\right)^{2n}\;
g^{-\frac{n}{2}}\;\leq\;
A_{2n}\;
g^{-\frac{n}{2}},
\nonumber\\[-1mm]
\label{lin.26}}
implies that
\BEG{align}{
&
\exp\left(
   -\int\MRM{d}\mu_{(1-t)\,\gamma}(\xi)\;
   V(\phi+\xi,g)
\right)\;
\frac{\gamma}{2}
\left(
   \PAR{\phi}
   \int\MRM{d}\mu_{(1-t)\gamma}(\xi)\;
   V(\phi+\xi,g)
\right)^{2}
\nonumber\\&\leq\;
\exp\left(
   -\frac{g\,\phi^{4}}{2}+15\,\upsilon\,g
\right)\;
8\,\gamma\,g^{2}\;
\biggl(
   \phi^{6}+9\,\upsilon^{2}\,\phi^{2}
\biggr)
\nonumber\\&\leq\;
\exp\left(
   -\frac{g\,\phi^{4}}{4}+15\,\upsilon\,g
\right)\;
8\,\gamma\,g^{\frac{1}{2}}\;
\biggl(
   A_{6}+9\,g\,\upsilon^{2}\,A_{2}
\biggr).
\quad\Box
\label{lin.27}}
Eq.~\REF{lin.26} shows that each $\phi$ in the downstairs 
factor kills $g^{\frac{1}{4}}$. Therefore, each of the two 
$\phi$-derivatives in its calculation yields $g^{\frac{1}{4}}$. 
Two $\phi$-derivatives give a total factor of $g^{\frac{1}{2}}$. 

\subsubsection{Fluctuation integral}

Half of the stability estimate has now been used up for the 
control of the downstairs factor. The other half suffices to
do the fluctuation integral.
\BEG{LEM}{Let $b$ and $c$ be positive constants. For all 
$\psi\in\MBB{R}$, we have that
\BEG{gather}{
\int\MRM{d}\mu_{\gamma}(\zeta)\;
\MRM{e}^{-b\,(\psi+\zeta)^{4}}\;\leq\;
\exp\left(
   -\frac{b\,c}{1+b\,c\,\gamma}\;\frac{\psi^{2}}{2}
   +\frac{b\,c^{2}}{16}
\right).
\label{lin.28}}
\label{err.5}}
{\bf Proof}\hspace{1mm}
From \REF{lin.6}, we deduce that
\BEG{gather}{
b\,\phi^{4}\;\geq\;
\frac{b\,c}{2}\phi^{2}-\frac{b\,c^{2}}{16}.
\label{lin.29}}
The Gaussian convolution of a Gauss function is again a 
Gauss function. From its explicit form, we find \REF{lin.28}.
$\Box$
\BEG{LEM}{For all $(\phi,g)\in\MBB{R}\times\MBB{G}$, we have 
that
\BEG{gather}{
\int\MRM{d}\mu_{t\,\gamma}(\zeta)\;
\MRM{e}^{
   -\int\MRM{d}\mu_{(1-t)\,\gamma}(\xi)\;
   V(\phi+\zeta+\xi)
}\;
\frac{\gamma}{2}\;
\left(
   \int\MRM{d}\mu_{(1-t)\,\gamma}(\xi)\;
   V(\phi+\zeta+\xi)
\right)^{2}\;\leq\;
\nonumber\\
C(g)\;\sqrt{g}\;
\exp\biggl(
   -\frac{1}{2}\,
   \frac{g^{\rho}}{1+g^{\rho}\,\gamma}\,
   \phi^{2}
   +15\,\upsilon^{2}\,g
   +\frac{1}{4}\,g^{2\rho-1}
\biggr).
\label{lin.30}}
\label{err.6}}
{\bf Proof}\hspace{1mm}
From \REF{lin.22} and \REF{lin.28}, with $b\,=\,\frac{g}{4}$ and 
$c\,=\,4\,g^{\rho-1}$, it follows that
\BEG{align}{
&\int\MRM{d}\mu_{t\,\gamma}(\zeta)\;
\MRM{e}^{
   -\int\MRM{d}\mu_{(1-t)\,\gamma}(\xi)\;
   V(\phi+\zeta+\xi)
}\;
\frac{\gamma}{2}\;
\left(
   \int\MRM{d}\mu_{(1-t)\,\gamma}(\xi)\;
   V(\phi+\zeta+\xi)
\right)^{2}
\nonumber\\&\leq\;
\int\MRM{d}\mu_{t\,\gamma}(\zeta)\;
C(g)\;\sqrt{g}\;
\exp\biggl(
   -\frac{g}{4}(\phi+\zeta)^{4}+15\,\upsilon^{2}\,g
\biggr)
\nonumber\\&\leq
C(g)\;\sqrt{g}\;
\exp\biggl(
   -\frac{1}{2}\,
   \frac{g^{\rho}}{1+g^{\rho}\,\gamma}\,
   \phi^{2}
   +15\,\upsilon^{2}\,g
   +\frac{1}{4}\,g^{2\rho-1}
\biggr).
\quad
\Box
\label{lin.31}}
In eq.~\REF{lin.14}, we also encounter another fluctuation 
integrals without downstairs factors. It is estimated in the
same manner. 
\BEG{LEM}{For all $(\phi,g)\in\MBB{R}\times\MBB{G}$, we have 
that
\BEG{gather}{
\int\MRM{d}\mu_{t\,\gamma}(\zeta)\;
\exp\left(
   -\int\MRM{d}\mu_{(1-t)\,\gamma}(\xi)\;
   V(\phi+\zeta+\xi)
\right)\;\leq\;
\nonumber\\ 
\exp\biggl(
   -\frac{1}{2}\,
   \frac{g^{\rho}}{1+g^{\rho}\,\gamma}\,
   \phi^{2}
   +15\,\upsilon^{2}\,g
   +\frac{1}{8}\,g^{2\rho-1}
\biggr).
\label{lin.32}}
\label{err.7}}
{\bf Proof}\hspace{1mm}
From \REF{lin.22} and \REF{lin.28}, with $b\,=\,\frac{g}{2}$ and 
$c\,=\,2\,g^{\rho-1}$, we find that
\BEG{align}{
&\int\MRM{d}\mu_{t\,\gamma}(\zeta)\;
\exp\biggl(
   -\int\MRM{d}\mu_{(1-t)\,\gamma}(\xi)\;
   V(\phi+\zeta+\xi)
\biggr)
\nonumber\\&\leq\;
\int\MRM{d}\mu_{t\,\gamma}(\zeta)\;
\exp\left(
   -\frac{g}{2}\phi^{4}+15\,\upsilon^{2}\,g
\right)
\nonumber\\&\leq\;
\exp\biggl(
   -\frac{1}{2}\,
   \frac{g^{\rho}}{1+g^{\rho}\,\gamma}\,
   \phi^{2}
   +15\,\upsilon^{2}\,g
   +\frac{1}{8}\,g^{2\rho-1}
\biggr).\quad\Box
\label{lin.33}}

\subsubsection{Scale transformation}

The remaining task is to combine \REF{lin.30} with \REF{lin.32}
and to rescale the field and the coupling. We insert these 
estimates into \REF{lin.14} to obtain the following error bound.
\BEG{LEM}{
Let $F_{1}$ be given by \REF{lin.9}.
For all $\epsilon\in\left(0,\frac{1}{2}\right)$ 
and all $L\in\{2,3,4,\ldots\}$ there exists a maximal coupling 
$g_{\MRM{max}}$, depending on $L$, $\epsilon$, 
such that for all $(\psi,g,t)\in\MBB{R}
\times\MBB{G}\times [0,1]$, we have that
\BEG{gather}{
\left\vert
   \PAR{t}
   X(\psi,g,t)
\right\vert
\;\leq\;
g^{\frac{1}{2}-\epsilon}\;F_{1}(g)\;
Z_{QU}(\phi,g).
\label{lin.34}}
\label{err.8}}
{\bf Proof}\hspace{1mm}
Insert \REF{lin.30} and \REF{lin.32} into \REF{lin.14} to conclude 
that
\BEG{align}{
\left\vert
   \PAR{t}X(\psi,g,t)
\right\vert&\leq\;
\alpha\;
(\delta\,g)^{\frac{1}{2}}\;
C(\delta\,g)\;
\exp\left(
   -\frac{\alpha}{2}\,
   \frac{(\delta\,g)^{\rho}}{1+(\delta\,g)^{\rho}}\,
   (\beta\,\psi)^{2}   
\right)
\nonumber\\&\phantom{\leq}\;
\exp\left(
   15\,\alpha\,\upsilon^{2}\,\delta\,g
   +\left(
      \frac{\alpha-1}{8}+\frac{1}{4}
   \right)\,
   (\delta\,g)^{2\,\rho-1}
\right).
\label{lin.35}}
We choose $g_{\MRM{max}}$ such that
\BEG{gather}{
\alpha\,\delta^{\frac{1}{2}}\,g^{\epsilon}\;
C(\delta\,g)\;
\exp\left\{
   15\,(\alpha\,\delta-1)\,\upsilon^{2}\,g
\right\}
\nonumber\\
\exp\left\{
      \left(
         \frac{\alpha-1}{8}+\frac{1}{4}
      \right)\,
      (\delta\,g)^{2\rho-1}
      -\frac{g^{2\rho-1}}{8}
\right\}\;
\exp\biggl\{
   -a_{QU}(g)
\biggr\}
\leq\;1
\label{lin.36}}
and
\BEG{gather}{
\alpha\,\beta^{2}\,
\frac{(\delta\,g)^{\rho}}{1+(\delta\,g)^{\rho}}\;\geq\;
b_{QU}(g).
\quad\Box
\label{lin.37}}
The condition \REF{lin.37} is easy to fulfill because
$\alpha\,\beta^{2}\,=\,L^{2}$ is on our side. The condition
\REF{lin.36} is also easy to fulfill, but it requires
$g_{\MRM{max}}$ to be exponentially small as a function on 
$L$. 
\BEG{LEM}{Let $\epsilon$, $L$, and $g_{\MRM{max}}$ be as 
in Lemma \ref{err.8}. Put 
\BEG{gather}{
F_{2}(g)\;=\;
g^{\frac{1}{2}-\epsilon}\;F_{1}(g).
\label{lin.38}}
Then $\| T_{1}(Z)\|_{F_{2}}\,\leq\,1$.
\label{err.9}}
Lemma \REF{err.8} is the first instant in this 
section where we need a small coupling argument. 
(Additionally, this $F_{1}$ limits the contraction mapping 
to small couplings.) To deal with large couplings case, 
we have to look for a modification of \REF{contract.9}. 
Since also \REF{lin.9} would require a modification, and 
since the exponent $\sigma\,=\,\frac{1}{2}-\epsilon$ is 
anyway too small to meet the condition \REF{all.4}, we 
will not eleborate on this possibility here. Instead, we will 
replace \REF{lin.1} by an approximant from higher 
order perturbation theory and modify the estimate of this 
section for this case.

\section{Perturbation theory in $g$ and $g^{2}\ln(g)$}

In this section, we first recall the formal power series solution
to the fixed point problem $T(V)\,=\,V$, where $T$ is defined by 
$S(Z)=\exp\bigl(-T(V)\bigr)$ with $Z=\exp(-V)$.  
As in \cite{Rolf/Wieczerkowski:1996}, we develop $V$ into a double 
perturbation expansion in both $g$ and $g^{2}\ln(g)$. 
We then prove a stability bound for the perturbative approximants
(of odd order in $g$), extending the analysis in 
\cite{Wieczerkowski:1997}.

\subsection{Formal power series representation}

In three dimensions, the renormalized $\phi^{4}$-trajectory 
is not expandable into a formal power series in $g$. However,
it does admit a formal power series representation in both
$g$ and $g^{2}\ln(g)$. See \cite{Rolf/Wieczerkowski:1996}.
To simplify the bookkeeping, we prefer $g$ and $\kappa=\ln(g)$ 
(instead of $g^{2}\ln(g)$) as formal expansion parameters. 
Let $V(\phi,g,\kappa)$ be given by a double formal power series
\BEG{gather}{
V(\phi,g)\;=\;
\sum_{r=1}^{\infty}\sum_{a=0}^{\left[\frac{r}{2}\right]}\;
V^{(r,a)}(\phi)\;g^{r}\;\kappa^{a}
\label{pert.1}}
with polynomial coefficients
\BEG{gather}{
V^{(r,a)}(\phi)\;=\;
\sum_{n=0}^{N(r,a)}\;
P_{2n}(\phi,\upsilon)\;
V^{(r,a)}_{2n},
\qquad
N(r,a)\;=\;r-2\,a+1.
\label{pert.2}}
The maximal number of fields $N$ at a given order $(r,a)$ is 
peculiar to $\phi^{4}$-theory. For safety reasons, we define
\BEG{gather}{
V^{(r,a)}_{2n}\,=\,0
\qquad
n\,>\,N(r,a).
\label{pert.3}}
Also, we set the order zero to zero. 
To first order, the trajectory is defined to be a pure normal 
ordered $\phi^{4}$-vertex,
\BEG{gather}{
V^{(1,a)}_{2n}\;=\;
\delta_{a,0}\;\delta_{n,2}.
\label{pert.4}}
The perturbative fixed point turns out to have two free 
parameters one for each resonance. See 
\cite{Rolf/Wieczerkowski:1996}. All of these solutions are 
suitable approximants for the contraction mapping. 
We set both parameters to zero,
\BEG{gather}{
V^{(2,0)}_{2}\;=\;
V^{(3,0)}_{0}\;=\;0.
\label{pert.5}}
The choice \REF{pert.5} has the advantage is to have a minimal 
number of vertices. 

The formal power series \REF{pert.1} supplies us with a 
sequence of polynomial approximants
\BEG{equation}{
V^{(r_{\MRM{max}})}(\phi,g)\;=\;
\sum_{r=1}^{r_{\MRM{max}}}\sum_{a=0}^{\left[\frac{r}{2}\right]}\;
V^{(r,a)}(\phi)\;g^{r}\;\ln(g)^{a}
\label{poly.1}}
labeled by the maximal power $r_{\MRM{max}}$ of $g$. The first
of which, $r_{\MRM{max}}=1$, is the above linear approximant.
The default value of $r_{\MRM{max}}$ will be seven in the following.

\subsubsection{Recursion relation}

To be a fixed point of the extended renormalization group 
transformation $T$ (the transformation for the interaction
$V$) in the sense of a double formal power series, the 
coefficients have to satisfy the following recursion relation.
Let $\bigl\langle \MCA{O}_{1};\ldots;\MCA{O}_{n}
\bigr\rangle^{T}_{\gamma,\Phi}$ denote the cumulants 
associated with the Gaussian moments
\BEG{equation}{
\bigl\langle
   \MCA{O}_{1}\cdots\MCA{O}_{n}
\bigr\rangle_{\gamma,\Phi}\;=\;
\int\MRM{d}\mu_{\gamma}(\zeta)\;
\MCA{O}_{1}(\Phi+\zeta)\cdots
\MCA{O}_{n}(\Phi+\zeta).
\label{cumulants}}
\BEG{LEM}{Let $V$ be given by \REF{pert.1}. 
Let $K(V)^{(r,a)}_{2n}$ be the coefficients defined 
by\footnote{The Gaussian integral projects onto the 
$P_{2n}(\psi)$-component of the cumulant.}
\BEG{gather}{
K(V)^{(r,a)}_{2n}\;=\;
\nonumber\\
\alpha\,\delta^{r}\;
\sum_{i=2}^{r}\frac{(-1)^{i+1}}{i!}\;
\sum_{r_{1}=1}^{r}\sum_{a_{1}=0}^{\left[\frac{r_{1}}{2}\right]}
\cdots
\sum_{r_{i}=1}^{r}\sum_{a_{i}=0}^{\left[\frac{r_{i}}{2}\right]}\;
\delta_{r,r_{1}+\cdots+r_{i}}\;\delta_{a,a_{1}+\cdots+a_{i}}\;
\nonumber\\ \times
\frac{1}{(2\,n)!\,\upsilon^{n}}\;
\int\MRM{d}\mu_{\upsilon}(\psi)\;
P_{2n}(\psi,\upsilon)\;
\biggl\langle
   V^{(r_{1},a_{1})};\cdots;V^{(r_{i},a_{i})}
\biggr\rangle^{T}_{\gamma,\beta\psi}.
\label{pert.6}}
Then $Z\,=\,\MRM{e}^{-V}$ is a fixed point of $S$ in the sense
of a formal double power series if and only if 
\BEG{gather}{
\left(
   1-L^{3-n-r}
\right)\;
V^{(r,a)}_{2n}\;=\;
\nonumber\\
-L^{-r}\;
\sum_{b=a+1}^{\left[\frac{r}{2}\right]}
\binom{b}{a}\;
\ln(L)^{b-a}\;
V^{(r,b)}_{2n}
+K(V)^{(r,a)}_{2n}.
\label{pert.7}}
}
To derive this set of equations, one performs a cumulant 
expansion for the hierarchical renormalization group, 
rescales the coupling, and compares equal double orders 
$(r,a)$.
\BEG{LEM}{The system of equations \REF{pert.7} has a unique 
solution of the form \REF{pert.1} with the properties 
\REF{pert.4} and \REF{pert.5}. 
}
{\bf Proof}\hspace{1mm}
The set of equations \REF{pert.7} can be solved recursively.
The condition \REF{pert.3} iterates through the 
recursion.\footnote{One cannot build connected diagrams with more 
than $2(r-2\,a+1)$ external legs from $r-2\,a$ vertices 
$g\,:\phi^{4}:_{\upsilon}$ and $a$ vertices $g^{2}\ln(g)\,
:\phi^{2}:_{\upsilon}$.}
One proceeds forwards in the order $r-1\rightarrow r$ and, at 
the order $r$, backwards in $a\rightarrow a-1$. Suppose that 
we have computed $V^{(s,b)}_{2n}$ both 
\BEG{enumerate}{
\item for all $(s,b)$ with $1\,\leq\,s\,\leq\,r-1$ and 
$0\,\leq\,b\leq\left[\frac{s}{2}\right]$ and
\item for all $(s,b)$ with $s\,=\,r$ and 
$a+1\,\leq\,\,b\leq\,\frac{r}{2}$.
}
Then this data determines the right hand side of \REF{pert.7}.
Therefrom, we compute $V^{(r,a)}_{2n}$ for all $n\,\leq\,N(r,a)$. 
We find two cases.
\BEG{enumerate}{
\item Non-resonant case: If $3-n-r\,\neq\,0$, then \REF{pert.7}
   determines $V^{(r,a)}_{2n}$.
\item Resonant case: If $(r,n)\in\{(2,1),(3,0)\}$, then the 
   left hand side of \REF{pert.7} is zero. In both cases, we
   find a constraint on the right hand side of \REF{pert.7}.
} 
The two resonances are resolved by logarithmic 
corrections. Consider the mass resonance $(2,1)$. Since\footnote{ 
$g^{2}\,\ln(g)\,:\phi^{2}:_{\upsilon}$ is not generated in the  
contraction of two vertices $g\,:\phi^{4}:_{\upsilon}$.} 
\BEG{gather}{
K(V)^{(2,1)}_{2}\;=\;0,
\label{pert.8}}
the equation labeled by $(r,a,n)\,=\,(2,1,1)$ is automatically 
satisfied. The equation with $(r,a,n)\,=\,(2,0,1)$ becomes
\BEG{gather}{
0\;=\;
-L^{-2}\;\ln(L)\;V^{(2,1)}_{2}
+K(V)^{(2,0)}_{2}.
\label{pert.9}}
We use it to determine $V^{(2,1)}_{2}$. The other parameter 
$V^{(2,1)}_{0}$ is unconstrained. We put it to zero. The 
vacuum resonance $(3,0)$ is analogously resolved. $\Box$

\subsection{Stability bound}

To any finite order, perturbation theory furnishes approximate
solutions to our fixed point problem, which are polynomials
in $\phi$ with coefficients that are polynomials in both 
$g$ and $g^{2}\ln(g)$. We intend to a polynomial of this kind 
as the approximate fixed point in our contraction mapping. 
To this aim, we need to prove two properties, a stability bound 
and an error bound. Both bounds will be proved analogously to 
\cite{Wieczerkowski:1997}. 
  
\subsubsection{Tree approximation}

We first prove stability for the tree approximation, which is 
defined as the polynomial in $\phi$, whose coefficients are 
simplified to their leading powers in $g$.
This bound extends by continuity to a stability bound 
for the complete perturbative approximant in a small coupling 
region. 

The set of coefficients $V_{tree}^{(r)}\,=\,V^{(r,0)}_{2(r+1)}$ 
can be computed independently of the others. They define a
tree approximation
\BEG{gather}{
V_{tree}(\phi,g)\;=\;
\sum_{r=1}^{\infty}\;
\phi^{2(r+1)}\;g^{r}\;V_{tree}^{(r)}
\label{pert.10}} 
to the renormalized $\phi^{4}$-trajectory\footnote{The sum of 
tree graph contributions is in fact convergent. We will not 
use this fact, since we are dealing with finite order 
approximants, which are polynomials in $\phi$.}. Notice that we have
replaced $:\phi^{2n}:_{\upsilon}$ by its highest term $\phi^{2n}$.
\BEG{LEM}{The recursion relation for $V^{(r)}_{tree}$
decouples. It reads
\BEG{gather}{
\left(1-L^{2(1-r)}\right)\;V^{(r)}_{tree}\;=\;
\nonumber\\
L^{3-r}\;
\sum_{i=2}^{r}\frac{(-1)^{i+1}}{i!}
\sum_{r_{1}=1}^{r}\cdots\sum_{r_{i}=1}^{r}\;
\delta_{r,r_{1}+\cdots+r_{i}}\;
V_{tree}^{(r_{1})}\cdots V_{tree}^{(r_{i})}
\nonumber\\ \times
\frac{1}{(2(r+1))!\,\upsilon^{2(r+1)}}\;
\int\MRM{d}\mu_{\upsilon}(\psi)\;
P_{2(r+1)}(\psi,\upsilon)\;
\biggl\langle
   P_{2(r_{1}+1)};\cdots;P_{2(r_{i}+1)}
\biggr\rangle^{T}_{\gamma,\beta\psi}\;
\nonumber\\[-1mm]
\label{pert.11}}
} 
All vertices in the tree approximation are irrelevant in the
extended powercounting. In particular, there are no resonances 
in the tree approximation. The tree coefficients have a 
simple sign pattern.
\BEG{LEM}{For all $r\geq 1$, $V_{tree}^{(r)}\,=\,(-1)^{r+1}\,
\left\vert V_{tree}^{(r)}\right\vert$.}
{\bf Proof}\hspace{1mm}
An induction on the order $r$. $\Box$\\[2mm]
It follows that all tree approximants with even maximal order 
$r_{\MRM{max}}$ are unstable. Therefore, we will restrict our
attention to the friendly approximants with odd maximal order.

The sign pattern remains valid at sufficiently small couplings.
To realize this, assemble the perturbative approximant with loop 
contributions.
\BEG{LEM}{Let $V_{tree}^{(-1)}\,=\,V_{tree}^{(0)}\,=\,0$. Then 
we have that
\BEG{gather}{
\sum_{r=1}^{r_{\MRM{max}}}\sum_{a=0}^{\left[\frac{r}{2}\right]}
V^{(r,a)}(\phi)\;g^{r}\;\ln(g)^{a}\;=\;
\sum_{n=0}^{r_{\MRM{max}}+1}
\phi^{2\,n}\;
g^{n-1}\;
\biggl(
   V^{(n-1)}_{tree}
   +\lambda_{2n}(g)
\biggr)
\nonumber\\[-1mm]
\label{pert.12}}
with $\lambda_{2n}(g)\,=\,O(g,g^{2}\,\ln(g))$.
}
The tree coefficients are the leading terms of the perturbative
vertex functions at small couplings. The loop corrections are 
continuous functions of $g$ (since they are polynomials in 
$g$ and $g^{2}\ln(g)$). Therefore, properties like the sign 
pattern at $g=0$ extend to a finite region $g\in [0,g_{\MRM{max}}]$ 
of small couplings.

\subsubsection{Effective $\phi^{4}$-coupling}

All $g^{2}\ln(g)$-terms are subleading. For this reason, the tree 
graph bound of \cite{Wieczerkowski:1997} applies also to the three 
dimensional model. 

Let $\mu_{2n}(g)\,=\,V^{(n-1)}_{tree}+\lambda_{2n}(g)$ (a 
polynomial in $g$ and $g^{2}\ln(g)$) so that the perturbative
approximant becomes
\BEG{equation}{
V^{(r_{\MRM{max}})}(\phi,g)\;=\;
\sum_{n=0}^{r_{\MRM{max}}+1}\phi^{2n}\;g^{n-1}\;\mu_{2n}(g).
\label{eff.1}}
\BEG{LEM}{
For all $r_{\MRM{max}}\geq 1$, there exists a 
maximal coupling $g_{\MRM{max}}>0$ 
(depending on $r_{\MRM{max}}$)  
such that for all $g\in\MBB{G}$ and 
$n\in\{2,3,\ldots,r_{\MRM{max}}+1\}$,
\BEG{equation}{
\mu_{2n}(g)\;=\;
(-1)^{n}\;
\vert\mu_{2n}(g)\vert.
\label{eff.2}}
}
In the following, we will assume that $g_{\MRM{max}}$ is 
sufficiently small such that \REF{eff.2} holds.

The following statements are presumably true for any finite 
order $r_{\MRM{max}}\in 2\MBB{N}+1$. As a part of their proofs,
we will have to compute certain coefficients recursively. 
I have only done this up to the (already ridiculously high) 
order $r_{\MRM{max}}=99$.  
\BEG{LEM}{
Let $r_{\MRM{max}}\in\{1,3,5,\ldots,99\}$. 
Then there exists a maximal coupling $g_{\MRM{max}}>0$ such 
that, for all $g\in\MBB{G}$, \REF{eff.1} is bounded from below by 
\BEG{equation}{
V^{(r_{\MRM{max}})}(\phi,g)\;\geq\;
\sum_{n=0}^{1}\phi^{2n}\;g^{n-1}\;\mu_{2n}(g)
+\phi^{4}\;g\;\rho_{4}(g),
\label{eff.3}}
where $\rho_{4}(g)$ is the solution to the recursion relation
\BEG{equation}{
\rho_{4n}(g)\;=\;
\mu_{4n}(g)
-\frac{\mu_{4n+2}(g)^{2}}{4\,\rho_{4n+4}(g)}
\label{eff.4}}
with the initial condition
\BEG{equation}{
\rho_{2(r_{\MRM{max}}+1)}(g)\;=\;
\mu_{2(r_{\MRM{max}}+1)}(g).
\label{eff.5}}
}
{\bf Proof}\hspace{1mm}
The proof is an induction on powers of $\phi^{4}$. (Notice 
that $2(r_{\MRM{max}}+1)\in 4\MBB{N}$.) The induction
step follows from 
\BEG{gather}{
\phi^{4n}\;g^{2n-1}\;
\biggl\{
   \mu_{4n}(g)
   +\phi^{2}\,g\,\mu_{4n+2}(g)
   +\phi^{4}\,g^{2}\,\rho_{4n+4}(g)
\biggr\}
\nonumber\\=\;
\phi^{4n}\;g^{2n-1}\;
\left\{
   \mu_{4n}(g)
   -\frac{\mu_{4n+2}(g)^{2}}{4\,\rho_{4n+4}(g)}
   +\rho_{4n+4}(g)\,
   \left(
      \phi^{2}\,g
      +\frac{\mu_{4n+2}}{2\,\rho_{4n+4}}
   \right)^{2}
\right\}
\nonumber\\[-1mm]
\label{eff.6}}
since $\rho_{4n+4}(g)$ is positive for small couplings. 
$\Box$\\[2mm]
A proof of the positivity of the effective $\phi^{4n}$-couplings
is given below. 
The solution of the recursion relation \REF{eff.4} is a rational
function 
\BEG{equation}{
\rho_{4}(g)\;=\;
\frac{P(g,g^{2}\ln(g))}{Q(g,g^{2}\ln(g))},
\label{eff.7}}
where $P$ and $Q$ are polynomials in $g$ and $g^{2}\ln(g)$. 
\BEG{LEM}{
Let $r_{\MRM{max}}\in\{1,3,5,\ldots,99\}$. There exists a positive 
number $c\,>\,0$ and a maximal coupling $g_{\MRM{max}}\,>\,0$ 
such that for all $g\in\MBB{G}$, we have that
\BEG{equation}{
\rho_{4}(g)\;\geq\;
c.
\label{eff.8}} 
}
{\bf Proof}\hspace{1mm}
The value $\rho_{4}(0)$ is determined as the solution of the 
recursion relation
\BEG{equation}{
\rho_{4n}(0)\;=\;
V^{(2n-1)}_{tree}
-\frac{V^{(2n)}_{tree}}{4\,\rho_{4n+4}(0)},
\label{eff.9}}
with the initial condition 
$\rho_{2(r_{\MRM{max}}+1)}(0)\,=\,V^{(r_{\MRM{max}})}_{tree}$. 
Therefore, it depends only on the tree graph coefficients. 
An explicit computation shows that $\rho_{4}(0)$ is a positive number. 
Since \REF{eff.7} is a continuous function of $g$, the assertion 
follows. $\Box$\\[2mm]
The effective $\phi^{4n}$-couplings at order seven will be listed
below. In particular, the value of $\rho_{4}(0)$ at order seven is
\BEG{equation}{
\rho_{4}(0)\;=\;
\frac{4306}{5627}.
\label{eff.10}}
As a side remark, we mention that the effective $\phi^{4}$-coupling 
in the tree approximation is not a small number at large orders. 
(It presumably converges as the order is taken to infinity.)
\BEG{LEM}{Let $r_{\MRM{max}}\in\{1,3,5,\ldots,99\}$. 
There exist positive numbers 
$g_{\MRM{max}}$, $c$, and $a$ (all strictly larger than 
zero) such that for all $(\phi,g)\in\MBB{R}\times\MBB{G}$, we have 
that
\BEG{equation}{
V^{(r_{\MRM{max}})}(\phi,g)\;\geq\;
g\;
\left(
   \frac{c}{2}\,\phi^{4}
   -a
\right).
\label{eff.11}}
\label{highstability}}
{\bf Proof}\hspace{1mm}
From \REF{eff.3} and \REF{eff.5}, it follows that there exist 
two polynomials $A$ and $B$ and a positive number $c$ such that
\BEG{equation}{
V^{(r_{\MRM{max}})}(\phi,g)\;\geq\;
g\;
\biggl\{
   A\bigl(g,g^{2}\ln(g)\bigr)
   +B\bigl(g,g^{2}\ln(g)\bigr)\,\phi^{2}
   +c\,\phi^{4}
\biggr\}.
\label{eff.12}}
for all $\phi\in\MBB{R}$ and $g\in[0,g_{\MRM{max}}]$, where 
$g_{\MRM{max}}$ is a certain positive number. For this maximal
coupling, define $\|A\|_{\infty}=\sup_{g\in\MBB{G}}
\vert A(g,g^{2}\ln(g)\vert$ and analogously $\|B\|_{\infty}$.
Then we have that
\BEG{align}{
V^{(r_{\MRM{max}})}(\phi,g)\;&\geq\;
g\;
\biggl\{
   -\|A\|_{\infty}
   -\|B\|_{\infty}\,\phi^{2}
   +c\,\phi^{4}
\biggr\}
\nonumber\\&\geq\;
g\;
\biggl\{
   -\|A\|_{\infty}
   -\frac{\|B\|_{\infty}^{2}}{2\,c}
   +\frac{c}{2}\,\phi^{4}
\biggr\}.
\label{eff.13}} 
Put $a\,=\,\|A\|_{\infty}+\frac{\|B\|_{\infty}^{2}}{2\,c}$
to obtain the assertion. $\Box$\\[2mm]
The remaining stability analysis is completely analogous to 
the linear case.
\BEG{LEM}{
Let $r_{\MRM{max}}\in\{1,3,5,\ldots,99\}$. 
Let $g_{\MRM{max}}$ be as in Lemma \ref{highstability}.
Let $F_{1}:\MBB{G}\rightarrow\MBB{R}^{+}$ be given by
\BEG{equation}{
F_{1}(g)\;=\;
\exp\left(
   a\,g+\frac{1}{8\,c}\,g^{2\rho-1}
\right).
\label{eff.14}} 
Then $Z^{(r_{\MRM{max}})}\,=\,\exp\bigl\{-V^{(r_{\MRM{max}})}\bigr\}$ 
is bounded 
in the norm associated with $F_{1}$. We have that
\BEG{equation}{
\|Z^{(r_{\MRM{max}})}\|_{F_{1}}\;\leq\;1.
\label{eff.15}}
}
This shows that the perturbative approximants are indeed in the 
domain of the extended renormalization group.
The stability bound is complete aside of a proof of the positivity
of $\rho_{4n+4}(g)$. By continuity, it suffices to prove the 
positivity of $\rho_{4n+4}(0)$, which depends only on the tree
coefficients. 

\subsubsection{Computation of tree coefficients}

There is another way to compute the tree coefficients than by
the recursion relation \REF{pert.11}, which uses a Hamilton-Jacobi
differential equation. This other way is both simpler than to 
iterate \REF{pert.11} and it also relies on an interpolation 
formula, which we will need independently in the error 
bound.

The perturbative renormalization group is the formal power series
solution of the non-linear transformation
\BEG{equation}{
T(V)(\psi,g)\;=\;
-\alpha\;
\ln\biggl[
   \int\MRM{d}\mu_{\gamma}(\zeta)\;
   \exp\bigl\{
      -V(\beta\psi+\zeta,\delta g)
   \bigr\}
\biggr].
\label{tree.1}}
It can be computed in two steps. Step one is the fluctuation 
integral
\BEG{equation}{
W(\phi,g,t)\;=\;
-\ln\biggl[
   \int\MRM{d}\mu_{t\gamma}(\zeta)\;
   \exp\bigl\{
      -V(\phi+\zeta,\delta g)
   \bigr\}
\biggr],
\label{tree.2}}
evaluated at $t=1$. The interpolated quantity satisfies the
renormalization group differential equation
\BEG{equation}{
\PAR{t}W(\phi,g,t)\;=\;
\frac{\gamma}{2}\;
\left[
   \frac{\partial^{2}}{\partial\phi^{2}}W(\phi,g,t)
   -\left\{
      \PAR{\phi}W(\phi,g,t)
   \right\}^{2}
\right]
\label{tree.3}}
in the sense of a formal powerseries, with the inital condition
\BEG{equation}{
W(\phi,g,0)\;=\;
V(\phi,g).
\label{tree.4}}
Step two is the scale transformation of the result of step one,
\BEG{equation}{
T(V)(\psi,g)\;=\;
\alpha\;
W\bigl(
   \beta\psi,\delta g,1
\bigr).
\label{tree.5}}
Consider the tree approximation hereof. The tree approximation
affects only step one. Eq.~\REF{tree.3} has to be replaced by the 
Hamilton-Jacobi equation
\BEG{equation}{
\PAR{t}W_{tree}(\phi,g,t)\;=\;
-\frac{\gamma}{2}\;
\left\{
   \PAR{\phi}W_{tree}(\phi,g,t)
\right\}^{2}
\label{tree.6}}
with the initial condition 
\BEG{equation}{
W_{tree}(\phi,g,0)\;=\;
V_{tree}(\phi,g).
\label{tree.7}}
The condition of renormalization invariance becomes
\BEG{equation}{
V_{tree}(\phi,g)\;=\;
\alpha\;
W_{tree}\bigl(
   \beta\psi,\delta g,1
\bigr).
\label{tree.8}}
\BEG{LEM}{
The Hamilton-Jacobi equation \REF{tree.6} has a unique formal 
power series solution 
\BEG{equation}{
W(\phi,g,t)\;=\;
\sum_{n=2}^{\infty}
B_{2n}(t)\;\phi^{2n}\,g^{n-1}
\label{tree.9}}
with the boundary condition
\BEG{equation}{
W(\phi,g,1)\;=\;
\alpha^{-1}\;
W\bigl(
   \beta^{-1}\phi,\delta^{-1}g
\bigr)
\label{tree.10}}
and $B_{4}(0)=1$. It reads 
\BEG{equation}{
B_{2n}(t)\;=\;
b_{2n}\;
\left\{
   -\gamma\,
   \left(
      \frac{1}{L^{2}-1}
      +t
   \right)
\right\}^{n-2},
\label{tree.11}}
where the coefficients $b_{2n}$ are recursively determined by
\BEG{equation}{
(n-2)\,b_{2n}\;=\;
2\sum_{m+l=n+1}m\,l\,b_{2m}\,b_{2l}
\label{tree.12}}
with the initial condition $b_{4}=1$.
} 
We remark that this solution makes sense beyond a formal power 
series. To see this, one writes
\BEG{equation}{
W(\phi,g,t)\;=\;
g^{-1}\;
W\bigl(
   \sqrt{g}\,\phi,1,t
\bigr)
\label{tree.13}}
and shows inductively a bound on the positive coefficients 
$b_{2n}$. However, since we only need the formal power series 
solution, we leave this issue aside. 
\BEG{COR}{The tree graph coefficients are given by
\BEG{equation}{
V_{tree}^{(r)}\;=\;
b_{2(r+1)}\;
\left(
   \frac{\gamma}{1-L^{2}}
\right)^{r-1}.
\label{tree.14}}
}
This confirms the sign pattern of the tree coefficients. 

Once the tree coefficients are written on the blackboard, we 
proceed to compute the effective $\phi^{4n}$-couplings in 
the tree approximation. With the $t$-dependence switched on, 
their recursion relation reads
\BEG{equation}{
\rho_{4n}(t)\;=\;
B_{4n}(t)
-\frac{B_{4n+2}(t)^{2}}{\rho_{4n+4}(t)}
\label{tree.15}}
starting at 
\BEG{equation}{
\rho_{2(r_{\MRM{max}}+1)}(t)\;=\;
B_{2(r_{\MRM{max}}+1)}(t).
\label{three.16}}
\BEG{LEM}{
The effective $\phi^{4n}$-couplings are given by 
\BEG{equation}{
\rho_{4n}(t)\;=\;
r_{4n}\;
\left\{
   \gamma\,
   \left(
      \frac{1}{L^{2}-1}
      +t
   \right)
\right\}^{2(n-1)}
\label{tree.17}}
with ($t$-independent) numbers $r_{4n}$ determined recursively by
\BEG{equation}{
r_{4n}\;=\;
b_{4n}
-\frac{b_{4n+2}^{2}}{r_{4n+4}}
\label{tree.18}}
where 
\BEG{equation}{
r_{2(r_{\MRM{max}}+1)}\;=\;
b_{2(r_{\MRM{max}}+1)}.
\label{tree.19}}
}
Remarkably, the effective $\phi^{4}$-coupling comes out to be 
independent of the interpolation parameter $t$. To seventh order
we find the following numbers.
\begin{center}
\begin{tabular}{|l|l|}\hline
\multicolumn{2}{|c|}{Tree coefficients} \\ \hline\hline
$n$ & $(-1)^{n}\,b_{2n}$ \\ \hline
2 & 1 \\ \hline
3 & -8 \\ \hline
4 & 96 \\ \hline 
5 & -1408 \\ \hline
6 & 23296 \\ \hline
7 & -417792 \\ \hline
8 & 7938048 \\ \hline
\end{tabular}                                                    
\hspace{10mm}          
\begin{tabular}{|l|l|}\hline
\multicolumn{2}{|c|}{Seventh order} \\ \hline\hline
$n$ & $r_{4n}$ \\ \hline
4 & 7938048 \\ \hline       
3 & 33817/19 \\ \hline
2 & 90032/1321 \\ \hline
1 & 4306/5627 \\ \hline
\end{tabular}
\end{center}
The error bound is complete for the order seven approximant. As 
a sidedish, we find the following useful bound which is uniform in
the interpolation parameter.
\BEG{LEM}{
The tree approximant of order seven 
\BEG{equation}{
W^{(7)}_{tree}(\phi,g,t)\;=\;
\sum_{n=2}^{8} B_{2n}(t)\;\phi^{2n}\;g^{n-1}
\label{tree.20}}
satisfies for all  
$(\phi,g,t)\in\MBB{R}\times\MBB{R}^{+}\times [0,1]$
the lower bound
\BEG{equation}{
W^{(7)}_{tree}(\phi,g,t)\;\geq\;
\frac{4306}{5627}\;g\phi^{4}.
\label{tree.21}}
}
The recursion relations \REF{tree.12} and \REF{tree.18} can be 
solved by computer algebra. Their solution proves the positivity
of the tree approximation (at least) up to the order $99$.  

\subsection{Error bound}

To prove an error bound for the higher order approximants, we
proceed analogous to the linear case. The main tool is a
generalization of the interpolation formula \REF{lin.17}.
From perturbation theory, we have a polynomial
\BEG{equation}{
V^{(r_{\MRM{max}})}(\phi,g)\;=\;
\sum_{r=1}^{r_{\MRM{max}}}
\sum_{a=0}^{\left[\frac{r}{2}\right]}
V^{(r,a)}(\phi)\;
g^{r}\;\ln(g)^{a}
\label{ebound.1}}
which satisfies the scaling relation
\BEG{equation}{
V^{(r_{\MRM{max}})}(\psi,g)\;=\;
\alpha\;
\sum_{n=1}^{r_{\MRM{max}}}
\frac{(-1)^{n+1}}{n!}\;
\MCA{P}^{(r_{\MRM{max}})}
\left\langle
   \left[
      V^{(r_{\MRM{max}})}(\cdot,\delta g);
   \right]^{n}
\right\rangle^{T}_{\gamma,\beta\psi}.
\label{ebound.2}}
Here $\MCA{P}^{(r_{\MRM{max}})}$ denotes a projector
\BEG{equation}{
\MCA{P}^{(r_{\MRM{max}})}
\bigl(
   g^{r}\ln(g)^{a}
\bigr)\;=\;
\begin{cases}
   g^{r}\ln(g)^{a} & \text{if $r\leq r_{\MRM{max}}$ and} \\
   0               & \text{else}.
\end{cases}
\label{ebound.3}}
The truncated cumulant expansion in \REF{ebound.2} contains 
terms of higher order than $g^{r_{\MRM{max}}}$. These 
are projected out by means of $\MCA{P}^{(r_{\MRM{max}})}$.
\BEG{DEF}{
Let $W^{(r_{\MRM{max}})}:\MBB{R}\times\MBB{G}\times [0,1]$ 
be defined by
\BEG{equation}{
W^{(r_{\MRM{max}})}(\psi,g,t)\;=\;
\sum_{n=1}^{r_{\MRM{max}}}
\frac{(-1)^{n+1}}{n!}\;
\MCA{P}^{(r_{\MRM{max}})}
\left\langle
   \left[
      V^{(r_{\MRM{max}})}(\cdot,\delta g);
   \right]^{n}
\right\rangle^{T}_{t\gamma,\psi}.
\label{ebound.4}}
}
This interpolation is identical with the formal power series
solution of \REF{tree.2}, projected 
to $\MCA{P}^{(r_{\MRM{max}})}$. Its boundary values are
\BEG{equation}{
W^{(r_{\MRM{max}})}(\psi,g,0)\;=\;
V^{(r_{\MRM{max}})}(\psi,g)\;=\;
\alpha\;
W^{(r_{\MRM{max}})}(\beta\psi,\delta g,1).
\label{ebound.5}} 
We use it to define the following generalization of 
\REF{lin.11} (the case $r_{\MRM{max}}=1$).
\BEG{DEF}{
Let $X^{(r_{\MRM{max}})}:\MBB{R}\times\MBB{G}\times [0,1]$ 
be defined by
\BEG{equation}{
X^{(r_{\MRM{max}})}(\psi,g,t)\;=\;
\left\{
   \int\MRM{d}\mu_{t\gamma}(\zeta)\;
   \exp\biggl(
      -W^{(r_{\MRM{max}})}(\beta\psi+\zeta,\delta g,1-t)
   \biggr)
\right\}^{\alpha}
\label{ebound.6}}
}
To be well defined, eq.~\REF{ebound.6} calls for a stability
bound for the interpolation \REF{ebound.5}. Postpone this 
issue for a short while. Eq.~\REF{ebound.6} yields the 
following representation for the error term.
\BEG{LEM}{
Let $r_{\MRM{max}}\in\{1,3,5,\ldots,99\}$. 
Then \REF{ebound.6} is well 
defined for the perturbative approximant \REF{ebound.1}.
We have that
\BEG{equation}{
T_{1}\bigl(
   Z^{(r_{\MRM{max}})}
\bigr)(\psi,g)\;=\;
X^{(r_{\MRM{max}})}(\psi,g,1)-X^{(r_{\MRM{max}})}(\psi,g,0).
\label{ebound.7}}
}
The usefulness of this representation relies on the 
following differential formula.
\BEG{LEM}{
Let $r_{\MRM{max}}\in\{1,3,5,\ldots,99\}$. Then \REF{ebound.6} is 
continuously differentiable in $t\in (0,1)$. We have that
\BEG{gather}{
\PAR{t}
X^{(r_{\MRM{max}})}(\psi,g,t)\;=\;
\nonumber\\
\alpha\;\left\{
   \int\MRM{d}\mu_{t\gamma}(\zeta)\;
   \exp\biggl(
      -W^{(r_{\MRM{max}})}(\beta\psi+\zeta,\delta g,1-t)
   \biggr)
\right\}^{\alpha-1}
\nonumber\\ \times
\int\MRM{d}\mu_{t\gamma}(\zeta)\;
\exp\biggl(
   -W^{(r_{\MRM{max}})}(\beta\psi+\zeta,\delta g,1-t)
\biggr)
\nonumber\\ \times
\frac{\gamma}{2}\;
\left[
   1-\MCA{P}^{(r_{\MRM{max}})}
\right]\;
\left(
   \PAR{\zeta}
   W^{(r_{\MRM{max}})}(\beta\psi+\zeta,\delta g,1-t)
\right)^{2}.
\nonumber\\[-2mm]
\label{ebound.8}}
}
{\bf Proof}
\BEG{align}{
&\PAR{t}
\int\MRM{d}\mu_{t\gamma}(\zeta)\;
\exp\left[
   -\MCA{P}\,
   \left\{
      \ln\int\MRM{d}\mu_{(1-t)\gamma}(\xi)\,
      \exp\bigl(
         -V(\phi+\zeta+\xi,g)
      \bigr)
   \right\}
\right]
\nonumber\\&=\;
\int\MRM{d}\mu_{t\gamma}(\zeta)\;
\left(
   \PAR{t}
   +\frac{\gamma}{2}
   \frac{\partial^{2}}{\partial\zeta^{2}}
\right)
\exp\left[
   -\MCA{P}\,
   \biggl\{
      \hspace{10mm}
   \biggr\}
\right]
\nonumber\\&=\;
\int\MRM{d}\mu_{t\gamma}(\zeta)\;
\exp\left[
   -\MCA{P}\,
   \biggl\{
      \hspace{10mm}
   \biggr\}
\right]
\nonumber\\&\phantom{=}\;\times
\Biggl[
-\left(
   \PAR{t}
   +\frac{\gamma}{2}
   \frac{\partial^{2}}{\partial\zeta^{2}}
\right)
\MCA{P}\,
\biggl\{
   \hspace{10mm}
\biggr\}
+\frac{\gamma}{2}
\left(
   \PAR{\zeta}
   \MCA{P}\,
   \biggl\{
      \hspace{10mm}
   \biggr\}   
\right)^{2}
\Biggr]
\label{ebound.9}}
and
\BEG{align}{
\left(
   \PAR{t}
   +\frac{\gamma}{2}
   \frac{\partial^{2}}{\partial\zeta^{2}}
\right)\;
\MCA{P}\;
\biggl\{
   \hspace{10mm}
\biggr\}
\;&=\;
\MCA{P}\;
\left(
   \PAR{t}
   +\frac{\gamma}{2}
   \frac{\partial^{2}}{\partial\zeta^{2}}
\right)\;
\biggl\{
   \hspace{10mm}
\biggr\}
\nonumber\\&=\;
\MCA{P}\;
\frac{\gamma}{2}\;
\left(
   \PAR{\zeta}\;
   \biggl\{
      \hspace{10mm}
   \biggr\}
\right)^{2}
\nonumber\\&=\;
\frac{\gamma}{2}\;
\MCA{P}\;
\left(
   \PAR{\zeta}\;
   \MCA{P}\;
   \biggl\{
      \hspace{10mm}
   \biggr\}
\right)^{2}.\quad\Box
\label{ebound.10}}
The upstairs factor is understood as a synonym for the 
truncated cumulant expansion. It is a polynomial expression.  
The downstairs factor in \REF{ebound.9} is in particular 
a polynomial expression and \REF{ebound.10} is a manipulation
of a polynomial expression.

The important feature of \REF{ebound.8} is that in the 
downstairs factor all orders lower than $r_{\MRM{max}}$ are 
cancelled by the $t$-dependent upstairs factor. To be 
well defined for all values of the interpolation parameter,
\REF{ebound.6} requires an additional stability bound.
\BEG{LEM}{
Let $r_{\MRM{max}}\in\{1,3,5,\ldots,99\}$.
There exist positive numbers $g_{\MRM{max}}$, $c$, and $a$
(all dependent on $r_{\MRM{max}}$) such that the following
stability bound holds for all 
$(\phi,g,t)\in\MBB{R}\times\MBB{G}\times [0,1]$:
\BEG{equation}{
W^{(r_{\MRM{max}})}(\phi,g,t)\;\geq\;
g\;\left(
   \frac{c}{2}\phi^{4}-a
\right).
\label{ebound.11}}
}
{\bf Proof}\\[2mm]
Since $W^{(r_{\MRM{max}})}(\phi,g,0)=V^{(r_{\MRM{max}})}(\phi,g)$,
we know that the bound \REF{ebound.11} is valid at $t=0$. 
Furthermore, we have shown that the effective
$\phi^{4}$-coupling is independent of $t$ in the tree 
approximation. The assertion now follows from the uniform continuity 
of the effective $\phi^{4}$-coupling of the complete perturbative
approximants. 

As the result of a truncated cumulant expansion, we have that
\BEG{equation}{
W^{(r_{\MRM{max}})}(\phi,g,t)\;=\;
\sum_{n=0}^{r_{\MRM{max}}+1}
\phi^{2n}\;
g^{n-1}\;
\mu_{2n}(g,g^{2},t).
\label{ebound.12}}
For $n\geq 2$, each coupling $\mu_{2n}(g,g^{2},t)$ is the sum of a 
tree term and loop contributions
\BEG{equation}{
\mu_{2n}(g,g^{2},t)\;=\;
B_{2n}(t)+\lambda_{2n}(g,g^{2}\ln(g),t).
\label{ebound.13}}
The tree term is given by \REF{tree.11}. The loop contributions 
are higher order corrections
\BEG{equation}{
\lambda_{2n}(g,g^{2}\ln(g),t)\;=\;
O(g,g^{2}\ln(g))
\label{ebound.14}}
as they are polynomials in $g$ and $g^{2}\ln(g)$ whose coefficients
are polynomials in $t$. 

Consider the effective $\phi^{4}$-coupling $\rho_{4}(g,t)$ 
defined as above.\footnote{To be precise, we should consider 
the collections of all $\phi^{4n}$-couplings $\rho_{4n}(g,t)$
and repeat the following reasoning for all of them.} 
As it is a continued fraction of couplings \REF{ebound.13},
it is a rational function (of $g$, $g^{2}\ln(g)$, and $t$) 
on some rectangle $[0,g_{\MRM{max}}^{\prime}]\times [0,1]$.  
Since the tree approximation has this particular $t$-dependence,
we have that $\rho_{4}(0,t)=r_{4}$ for all $t\in [0,1]$ at $g=0$. 
Let $c=r_{4}/2$. By continuity, there exists a positive number 
$g_{\MRM{max}}$ (with $0<g_{\MRM{max}}\leq g_{\MRM{max}}^{\prime}$)
such that for all $(g,t)\in [0,g_{\MRM{max}}]\times [0,1]$, we 
have that
\BEG{equation}{
\rho_{4}(g,t)\geq c.
\label{ebound.15}}
Taking care of the constant and quadratic term in $\phi$ 
analogously to \REF{eff.12} and \REF{eff.13}, the assertion 
follows. $\Box$\\[2mm]
As in the linear case, the stability bound on the interpolated
interaction is independent of the interpolation parameter.
The remaining analysis is completely analogous to the linear case. 

\subsubsection{Large field domination}

The harvest of the higher order perturbation theory is a 
higher power of $g$ in the bound after dominating the 
large fields by part of the stability estimate. 
\BEG{LEM}{
Let $r_{\MRM{max}}\in\{1,3,5,\ldots,99\}$. 
For all $(\phi,g,t)\in\MBB{R}\times
\MBB{G}\times[0,1]$, we have that
\BEG{gather}{
\exp\biggl\{
   -W^{(rmax)}(\phi,g,t)
\biggr\}\;
\frac{\gamma}{2}\;
\left[
   1-\MCA{P}^{(r_{\MRM{max}})}
\right]\;
\left(
   \PAR{\phi}\;
   W^{(rmax)}(\phi,g,t)
\right)^{2}
\nonumber\\ \leq
C(g)\; 
g^{r_{\MRM{max}}/2}\;
\exp\left(
   -\frac{c}{4}\,g\,\phi^{4}
   +a\,g
\right)
\label{ebound.16}}
for some polynomial $C\in\MBB{R}^{+}[g,g^{2}\ln(g)]$ 
(with positive coefficients).
}
{\bf Proof}\\[2mm]
The downstairs factor is a polynomial of the form
\BEG{gather}{
\frac{\gamma}{2}\;
\left[
   1-\MCA{P}^{(r_{\MRM{max}})}
\right]\;
\left(
   \PAR{\phi}\;
   W^{(rmax)}(\phi,g,t)
\right)^{2}
\nonumber\\ =\;
\sum_{n=0}^{r_{\MRM{max}}+1}
   \phi^{2n}\;g^{r_{\max}+1}\;B_{2n}(g,g^{2}\ln(g),t)
\nonumber\\ \phantom{=}\;
+\sum_{n=r_{\MRM{max}}+2}^{2r_{\MRM{max}}+1}
   \phi^{2n}\;g^{n-1}\;B_{2n}(g,g^{2}\ln(g),t)
\label{ebound.17}}
with certain polynomials $B_{2n}$. Notice that the 
projector affected only the first sum in \REF{ebound.17}. 
Notice also that the highest power of $\phi$ is 
$2(2(r_{\MRM{max}}+1))-2$,
where $-2$ comes from the two $\phi$-derivatives. 
With the help of the stability bound, we find the 
upper bound
\BEG{gather}{
\left\vert
\exp\biggl\{
   -W^{(rmax)}(\phi,g,t)
\biggr\}\;
\frac{\gamma}{2}\;
\left[
   1-\MCA{P}^{(r_{\MRM{max}})}
\right]\;
\left(
   \PAR{\phi}\;
   W^{(rmax)}(\phi,g,t)
\right)^{2}
\right\vert
\nonumber\\ \leq
\exp\left(
   -\frac{c}{4}\,g\,\phi^{4}
   +g\,a   
\right)
\Biggl\{
   \sum_{n=0}^{r_{\MRM{max}}+1}
      g^{r_{\MRM{max}}-\frac{n}{2}+1}\;
      A_{2n}\;
      \vert B_{2n}(g,g^{2}\ln(g),t)\vert
\nonumber\\ 
   +\sum_{n=r_{\MRM{max}}+2}^{2r_{\MRM{max}}+1}
      g^{\frac{n}{2}-1}\;
      A_{2n}\;
      \vert B_{2n}(g,g^{2}\ln(g),t)\vert
\Biggr\}
\label{ebound.18}}
where
\BEG{equation}{
A_{2n}\;=\;
\sup_{\phi\in\MBB{R}}\;
\left\vert
\exp\left(
   -\frac{c}{4}\phi^{4}
\right)\;
\phi^{2n}
\right\vert.
\label{ebound.19}}
Expand the polynomials $B$ and take the supremum of 
$t\in [0,1]$ in each term to arrive at the bound
\REF{ebound.16}. $\Box$\\[2mm]

\subsubsection{Fluctuation integral and scale transformation}

The fluctuation integral and the scale transformation 
are now identical to the linear case aside of a minimal
cosmetic modification to include the constant $c$.
We therefore do not repeat them here and jump to the 
following conclusion.
\BEG{LEM}{
Let $r_{\MRM{max}}\in\{1,3,5,\ldots,99\}$. 
Let $F_{1}$ be the function \REF{eff.14} from the stability 
bound. For all 
$\epsilon\in \left(0,\frac{r_{\MRM{max}}}{2}\right)$
and all $L\in\{2,3,4,\ldots\}$, there exists a maximal 
coupling $g_{\MRM{max}}$ such that for $(\psi,g,t)\in
\MBB{R}\times\MBB{G}\times [0,1]$, we have that
\BEG{equation}{
\left\vert
   \PAR{t}X^{(r_{\MRM{max}})}(\psi,g,t)
\right\vert\;
\leq\;
g^{\frac{r_{\MRM{max}}}{2}-\epsilon}\;
F_{1}(g)\;
Z_{QU}(\psi,g).
\label{ebound.20}}
\label{crucial}}
The error bound is an immediate consequence hereof.
\BEG{COR}{
Let $\epsilon$, $L$, and $g_{\MRM{max}}$ be as in Lemma
\ref{crucial}. Put 
\BEG{equation}{
F_{2}(g)\;=\;
g^{\frac{r_{\MRM{max}}}{2}-\epsilon}\;
F_{1}(g).
\label{ebound.21}}
Then we have that $Z^{(r_{\MRM{max}})}=
\exp\bigl(-V^{(r_{\MRM{max}})}\bigr)$ satisfies the bound
\BEG{equation}{
\left\|
   T_{1}
   \left(
      Z^{r_{\MRM{max}}}
   \right)
\right\|\;
\leq\;1.
\label{ebound.22}}
}
We have computed $V^{(r_{\MRM{max}})}(g)$ as a polynomial 
approximant of a formal double expansion in $g$ and $g^{2}\ln(g)$.
Then we have shown that, for sufficiently small (but finite) 
couplings, $Z^{(r_{\MRM{max}})}=\exp\bigl(-V^{(r_{\MRM{max}})}\bigr)$ 
satisfies both   
\begin{enumerate}
\item the stability bound 
$\|Z^{(r_{\MRM{max}})}\|_{F_{1}}\leq 1$, where $F_{1}$ is 
a function of the form \REF{small.1}, and
\item the error bound 
$\|T_{1}(Z_{1})\|_{F_{2}}\leq 1$, where $F_{2}$ is given 
by \REF{all.3}, with exponent 
$\sigma=\frac{r_{\MRM{max}}}{2}-\epsilon$. 
\end{enumerate}
For $r_{\MRM{max}}\geq 7$ and $\epsilon$ not too large, all 
assumptions of the contraction mapping are satisfied. The 
construction is complete.

\section{Conclusions and outlook}

The iteration of the contraction mapping provides a 
convergent representation for the $\phi^{4}_{3}$-trajectory.
It can be used to study the properties of the fixed point 
$Z_{\star}(\phi,g)$. One important problem, which can be 
shown, but which will not be shown here, is that $Z_{\star}(\phi,g)$
is positive. A brief discussion of its positivity is 
contained in \cite{Wieczerkowski:1997}. Other questions about
$Z_{\star}(\phi,g)$ could also be studied in principal, 
for instance the summability of perturbation theory, and 
analyticity properties of its Borel transform. 

A very interesting question is the behavior of $Z_{\star}(\phi,g)$
at large couplings. Conceivably, $Z_{\star}(\phi,g)$ 
connects the trivial fixed point at $g=0$ with the 
non-trivial infrared fixed point at $g=\infty$. The 
contraction mapping is potentially capable of a construction,
which is uniform in the running coupling. But such an 
enterprise requires a better approximate fixed point
$Z_{1}(\phi,g)$ than the one from perturbation theory. 
It is conceivable that one could extend the approximants 
from \cite{Koch/Wittwer:1986,Koch/Wittwer:1988,
Koch/Wittwer:1991,Koch/Wittwer:1994} to achieve this aim. 

The underlying scheme of this paper is to compute a renormalized
trajectory as a renormalization invariant curve in the 
unstable manifold of a renormalization group fixed point. 
This scheme is certainly translatable to virtually every theory
treated so far with the renormalization group. In particular, 
all hierarchical models mentioned in the introduction can 
be handled that way. We hope to present an extension of this 
method to the framework of polymer expansions and full models
in future work. Another aspect of this theory is the question how 
traditionally computed renormalized actions converge to the
renormalized trajectory. In other words, what is the domain 
of attraction of this extended fixed point of an extended 
renormalization group. This question is related to the problem
of renormalization group improved actions and also to the 
question how to truncate a renormalization group such as 
to maintain control of the errors. We hope to make progress
on these and other questions in this context in future work.\\[10mm]
{\bf\Large Acknowledgements}\\[4mm]
I would like to thank Andreas Pordt, Peter Wittwer, and 
Frank Zielen for helpful discussions of the rigorous renormalization 
group. 
I would also like to thank Horst Kn\"orrer and Manfred Salmhofer for 
the organization of the conference on {\sl Rigorous Renormalization}
in Ascona, where this work was presented.

\tableofcontents
\end{document}